\documentclass[11pt,a4paper]{article}
\usepackage{jheppub}
\usepackage[T1]{fontenc}
\usepackage{amssymb}
\usepackage{mathtools}

\usepackage{hyperref}
\usepackage{epsfig}
\usepackage{amsmath,latexsym,amssymb}
\usepackage{graphicx}
\usepackage{braket}
\usepackage{ulem}

\usepackage{mathrsfs}
\def\be{\begin{equation}}
\def\ee{\end{equation}}
\def\ba{\begin{eqnarray}}
\def\ea{\end{eqnarray}}

\newcommand{\bz}{\bar{z}}

\newcommand{\comment}[1]{}

\def\fc#1#2{{\frac{#1}{#2}}}
\newcommand{\req}[1]{(\ref{#1})}

\newcommand{\eea}{\end{eqnarray}}
\def\lf{\left}
\def\ri{\right}

\author{
Romain Ruzziconi${}^{1}$, Stephan Stieberger${}^{2}$, Tomasz R.\ Taylor${}^{3,4}$,\, Bin Zhu${}^{5}$\\[0.5cm]
$^1${\it Mathematical Institute, University of Oxford,  Oxford, OX2 6GG, U.K. }\\
$^2${\it Max--Planck--Institut f\"{u}r Physik,	Werner--Heisenberg--Institut, 85748 Garching, Germany}\\
 $^3${\it Department of Physics,
  Northeastern University, Boston, MA 02115, USA}\\
  $^4${\it Faculty of Physics, University of Warsaw, ul. Pasteura 5, 02-093 Warsaw, Poland}\\
$^5${\it School of Mathematics and Maxwell Institute for Mathematical Sciences,\\ University of Edinburgh,
EH9 3FD, U.K. }\\[0.2cm]
}

\emailAdd{romain.ruzziconi@maths.ox.ac.uk}
\emailAdd{stephan.stieberger@mpp.mpg.de}
\emailAdd{taylor@neu.edu}
\emailAdd{bzhu@exseed.ed.ac.uk}

\title{Differential Equations for Carrollian Amplitudes}

\abstract{Differential equations are powerful tools in the study of correlation functions in conformal field theories (CFTs). Carrollian amplitudes behave as correlation functions of Carrollian CFT that holographically describes asymptotically flat spacetime. We  derive linear differential equations satisfied by Carrollian MHV gluon  and graviton amplitudes. We obtain non-distributional solutions  for both the gluon and graviton cases. We perform various consistency checks for these differential equations, including compatibility with conformal Carrollian symmetries.}

\gdef\@fpheader{}
\makeatother

\begin{document}
\maketitle

\section{Introduction}

Differential equations satisfied by correlation functions are essential elements in Conformal Field Theory (CFT).
Some famous examples are Belavin-Polyakov-Zamolodchikov (BPZ) equations \cite{Belavin:1984vu}, Knizhnik-Zamolodchikov (KZ) equations \cite{Knizhnik:1984nr} in two dimensions, and differential equations of conformal partial waves by Dolan and Osborn \cite{Dolan:2011dv} in higher dimensions.
These differential equations place powerful constraints on the correlation functions.
For example, the four-point correlators in minimal models and in the WZW model can be determined by solving differential equations subjected to physically sensible monodromy properties \cite{Belavin:1984vu,Knizhnik:1984nr,DiFrancesco:1997nk}.

Celestial amplitudes for massless particles computed from Mellin transforms of scattering amplitudes in momentum basis with respect to the light-cone energies behave as conformal correlators on the celestial sphere \cite{Pasterski:2016qvg,Pasterski:2017kqt,Pasterski:2017ylz}.
This was understood as a change of basis from momentum-eigenstate basis to boost-eigenstate basis \cite{Stieberger:2018onx,Arkani-Hamed:2020gyp}, and has been under intensive research in the context of celestial holography due to the fact that the boost-eigenstate basis makes the conformal properties of scattering amplitudes manifest \cite{Pasterski:2016qvg,Pasterski:2017kqt,Pasterski:2017ylz,Stieberger:2018onx,Arkani-Hamed:2020gyp,Donnay:2018neh,Fan:2019emx,Pate:2019mfs,Adamo:2019ipt,Guevara:2019ypd,Puhm:2019zbl,Pate:2019lpp,Fotopoulos:2019vac,Banerjee:2020kaa,Donnay:2020guq,Guevara:2021abz,Strominger:2021mtt,Himwich:2021dau}.
One might hope that some stardard CFT techniques, e.g., differential equations can be applied to the study of celestial amplitudes. This was initiated by \cite{Banerjee:2020zlg} for MHV graviton amplitudes and \cite{Banerjee:2020vnt} for MHV gluon amplitudes respectively. See also \cite{Banerjee:2021cly,Pasterski:2021fjn,Hu:2021lrx,Hu:2022bpa,Adamo:2022wjo,Banerjee:2023zip,Saha:2023hsl,Banerjee:2023bni} for further developments.

Recently, an alternative change of basis for massless particles has gained a lot of interest since \cite{Donnay:2022aba,Bagchi:2022emh,Donnay:2022wvx}, which is to perform Fourier transforms with respective to the energies rather than Mellin transforms.
In this way, one obtain Carrollian amplitudes which provide non-trivial putative data for a three-dimensional Carrollian CFT that has been proposed to describe four-dimensional asymptotically flat spacetime \cite{Donnay:2022aba,Bagchi:2022emh,Donnay:2022wvx}. Carrollian CFTs are field theories exhibiting conformal Carroll (or BMS) symmetries as spacetime symmetries, and can be constructed from standard CFTs by formally taking the speed of light to zero, see e.g. \cite{Barnich:2012rz,Bagchi:2019xfx,Henneaux:2021yzg,deBoer:2021jej,Chen:2023pqf,Alday:2024yyj}. We refer to \cite{Donnay:2022aba,Bagchi:2022emh,Donnay:2022wvx,Salzer:2023jqv,Saha:2023abr,Nguyen:2023miw,Mason:2023mti,Bagchi:2023cen,Liu:2024nfc,Have:2024dff,Stieberger:2024shv,Adamo:2024mqn,Alday:2024yyj} for some explicit examples of Carrollian amplitudes. 

It has been shown that celestial amplitudes and Carrollian amplitudes are closely related to each other \cite{Donnay:2022wvx,Mason:2023mti}.
A natural question is whether we can find differential equations satisfied by Carrollian amplitudes, similar to what we have seen for their celestial counterparts.
In this paper, we provide a definitive answer to this question and derive Carrollian differential equations at null infinity. We investigate the solution space of these equations and find non-distributional solutions beyond Carrollian MHV amplitudes. 

This paper is organized as follows.
In Section \ref{sec2}, we review some basics of Carrollian amplitudes. In Section \ref{sec3}, we review the differential equations satisfied by the celestial MHV gluon amplitudes and show how to translate them into the corresponding equations for the Carrollian MHV gluon amplitudes.
We find two non-distributional three-point solutions to the differential equations of Carrollian MHV gluon amplitudes. We discuss the higher-point case.
In Section \ref{sec4}, we review the differential equations satisfied by the celestial MHV graviton amplitudes and translate them into the corresponding equations for the Carrollian MHV graviton amplitudes.
In Section \ref{sec5}, we check explicitly that the differential equations satisfied by the Carrollian MHV amplitudes are consistent with conformal Carrollian symmetries at null infinity.   
Section \ref{sec6} discusses briefly some future directions. In Appendix \ref{secAppA}, we show an example of how to derive a differential equation for celestial MHV graviton amplitudes from a null state condition constructed from subleading and subsubleading graviton symmetries.

\section{Elements of Carrollian amplitudes} \label{sec2}
To compute Carrollian amplitudes from scattering amplitudes for massless particles in momentum basis, we parametrize the light-like momenta by
\be
p^{\mu} = \omega q^{\mu} =  \frac{1}{2}\, \omega(1+|z|^2, z+\bar{z}, -i (z-\bar{z}), 1-|z|^2) \, ,
\ee
where  $\omega$ is the light-cone energy. The scattering amplitudes in momentum basis can be expressed in terms of spinor products
\be
\langle ij\rangle= \sqrt{\omega_i\omega_j}\; z_{ij} \ \ ,\ \  \quad [ij] = -\sqrt{\omega_i\omega_j}\;\bar{z}_{ij}
\ee
and the usual scalar products
\be
s_{ij} = 2 p_ip_j = \omega_i\omega_j z_{ij}\bar{z}_{ij} \, .
\ee
The Carrollian amplitudes are obtained by performing Fourier transforms with respect to the light-cone energies:
\begin{equation}
\begin{split}
        C_n(\{u_1, z_1,\bar{z}_1\}^{\epsilon_1}_{J_1}, \dots,&\{u_n, z_n,\bar{z}_n\}^{\epsilon_n}_{J_n} ) \\
        &= \prod_{i=1}^n \left( \int_0^{+\infty} \frac{d\omega_i}{2\pi} e^{i\epsilon_i \omega_i u_i}\right)\mathcal{A}_n(\{\omega_1, z_1,\bar{z}_1\}^{\epsilon_1}_{J_1},\dots,\{\omega_n, z_n,\bar{z}_n\}^{\epsilon_n}_{J_n}) \, .  \label{eq:nCarrollian}
\end{split}
\end{equation}
where $\mathcal{A}_n$ are scattering amplitudes expressed in the usual momentum space basis, $\epsilon=\pm 1$ corresponds to outgoing $(+1)$ or incoming $(-1)$ particle and $J$ denotes the particle helicities. These Carrollian amplitudes can be re-interpreted as correlators in a Carrollian CFT at null infinity \cite{Donnay:2022wvx,Mason:2023mti}:
\begin{equation}
C_n(\{u_1, z_1,\bar{z}_1\}^{\epsilon_1}_{J_1}, \dots,\{u_n, z_n,\bar{z}_n\}^{\epsilon_n}_{J_n} ) = \langle \Phi^{\epsilon_1}_{J_1}(u_1,z_1,\bar{z}_1) \ldots \Phi^{\epsilon_n}_{J_n}(u_n,z_n,\bar{z}_n) \rangle
\label{holographic identification}
\end{equation} where $\Phi^{\epsilon_i}_{J_i}(u_i,z_i,\bar{z}_i)$ are conformal Carrollian primaries with weights 
\begin{equation}
    (k_i,\bar{k}_i) = \left(\frac{1+\epsilon_i J_i}{2},\frac{1-\epsilon_i J_i}{2} \right)
\label{carrollian weights}
\end{equation} 
The relation between the Carrollian amplitude $C_n$ and the corresponding celestial amplitude $\mathcal{M}_n$ is given by \cite{Donnay:2022aba,Donnay:2022wvx}
\begin{multline}
 \label{Btransfom2}  
   \mathcal{M}_n\left(\left\lbrace \Delta_1, z_1, \bar{z}_1\right\rbrace_{J_1}^{\epsilon_1}, \dots , \left\lbrace \Delta_n, z_n, \bar{z}_n\right\rbrace_{J_n}^{\epsilon_n}\right) \\ 
    =     \prod_{i=1}^n \left( \, (- i \epsilon_i)^{\Delta_i}\Gamma[\Delta_i] \int_{-\infty}^{+\infty} \frac{d u_i}{(u_i - i \epsilon_i \varepsilon)^{\Delta_i}} \right) \, C_n\left(\left\lbrace u_1, z_1, \bar{z}_1\right\rbrace_{J_1}^{\epsilon_1}, \dots , \left\lbrace u_n, z_n, \bar{z}_n\right\rbrace_{J_n}^{\epsilon_n}\right)
\end{multline} where $\varepsilon \to 0^+$ is a regulator. As explained in \cite{Donnay:2022wvx,Nguyen:2023miw,Mason:2023mti}, it is also useful to consider correlators of descendants by taking derivatives with respect to the $u$ coordinate. Indeed, $\partial_u$-descendants of conformal Carrollian primaries are also primaries. Adopting the notations of \cite{Mason:2023mti}, we have
\begin{align}
&C_n^{m_1\dots m_n}(\{u_1, z_1,\bar{z}_1\}^{\epsilon_1}_{J_1}, \dots,\{u_n, z_n,\bar{z}_n\}^{\epsilon_n}_{J_n}) = \partial_{u_1}^{m_1}\dots \partial_{u_n}^{m_n}C_n(\{u_1, z_1,\bar{z}_1\}^{\epsilon_1}_{J_1}, \dots,\{u_n, z_n,\bar{z}_n\}^{\epsilon_n}_{J_n}) \nonumber\\
&= \prod_{i=1}^n \left( \int_0^{+\infty} \frac{d\omega_i}{2\pi} \, (i\epsilon_i \omega_i)^{m_i} e^{i\epsilon_i \omega_i u_i}\right)\mathcal{A}_n(\{\omega_1, z_1,\bar{z}_1\}^{\epsilon_1}_{J_1},\dots,\{\omega_n, z_n,\bar{z}_n\}^{\epsilon_n}_{J_n}) \, , \label{eq:Cudescendants}
\end{align}
and a shorthand notation for $C_n^{1\dots1}$:
\be
\widetilde{C}_n(\{u_1, z_1,\bar{z}_1\}^{\epsilon_1}_{J_1}, \dots,\{u_n, z_n,\bar{z}_n\}^{\epsilon_n}_{J_n}) = C_n^{1\dots1}(\{u_1, z_1,\bar{z}_1\}^{\epsilon_1}_{J_1}, \dots,\{u_n, z_n,\bar{z}_n\}^{\epsilon_n}_{J_n}) \, . \label{eq:tildeC_n}
\ee
The explicit expressions of the $n$-point Carrollian MHV gluon and graviton amplitudes were derived in \cite{Mason:2023mti}.

%

\section{Differential equations for  Carrollian MHV gluon amplitudes} \label{sec3}

In this section, we derive the differential equations satisfied by Carrollian MHV gluon amplitudes, starting from their celestial version \cite{Banerjee:2020vnt,Hu:2021lrx}.
The differential equations satisfied by the full celestial MHV gluon amplitudes were obtain by Banerjee and Ghosh (BG) \cite{Banerjee:2020vnt}.
For the purpose of simplicity, we consider the coresponding differential equations satisfied by the color-ordered (partial) celestial MHV gluon amplitudes shown in \cite{Hu:2021lrx}.
The authors of \cite{Hu:2021lrx} also gave a simple explanation of the origin of BG equations in terms of BCFW shifts in momentum space.
For generic $n$-point color-ordered celestial MHV gluon amplitudes, they satisfy the following differential equations,
\begin{align}
&\left( \partial_i -\frac{\Delta_i}{z_{i-1,i}} -\frac{1}{z_{i+1,i}}\right)\mathcal{M}_n(1,\cdots,n) \nonumber\\
&+\left( \epsilon_i\epsilon_{i-1} \frac{\Delta_{i-1}-J_{i-1}-1+\bar{z}_{i-1,i}\bar{\partial}_{i-1}}{z_{i-1,i}}e^{\frac{\partial}{\partial\Delta_i}-\frac{\partial}{\partial\Delta_{i-1}}}\right) \mathcal{M}_n(1,\cdots,n)= 0 \, , \label{eq:nptBG}
\end{align}
where particle $i$ is a gluon with positive helicity. There is another set of equations that can be obtained from Eq.(\ref{eq:nptBG}) by $i-1 \leftrightarrow i+1$. $J_{i}$ is the helicity of gluon.

Notice that the operator $e^{\frac{\partial}{\partial\Delta}}$ in Eq.(\ref{eq:nptBG}) shifts the conformal dimension up by one. In momentum basis, this comes from having an extra $\omega$ in the integrand of the celestial amplitudes. Therefore, if we consider the corresponding Carrollian amplitudes, $e^{\frac{\partial}{\partial\Delta}}$ gets translated into $\frac{1}{i\epsilon}\partial_u$ acting on the Carrollian amplitudes:
\begin{align}
&\text{\ \ \ \ \ Celestial}  \qquad\qquad\qquad\qquad \text{\ \ \ Carrollian} \nonumber\\
&e^{\frac{\partial}{\partial\Delta_j}} \mathcal{M}_n(1,\cdots, n) \, \,  \longleftrightarrow \,\, \,  \frac{1}{i\epsilon_j}\partial_{u_j} C_n(1,\cdots,n) \, . \label{eq:etopartialDelta}
\end{align} This relation can also be directly deduced from the correspondence between Carrollian and celestial amplitudes in \eqref{Btransfom2}. 

The operator $e^{-\frac{\partial}{\partial\Delta}}$ in Eq.(\ref{eq:nptBG}) shifts the conformal dimension down by 1. Is is not obvious how to translate this operator into the Carrollian language, although one can formally write it as $\partial_{u}^{-1}$. To avoid this issue, we can simply shift the conformal dimension of $\Delta_{i-1}$ up by one in Eq.(\ref{eq:nptBG}) by choosing $\Delta_{i-1}\rightarrow\Delta_{i-1}+1$ in $\mathcal{M}_n$. Eq.(\ref{eq:nptBG}) becomes
\begin{align}
&\left( \partial_i -\frac{\Delta_i}{z_{i-1,i}} -\frac{1}{z_{i+1,i}}\right) e^{\frac{\partial}{\partial\Delta_{i-1}}}\mathcal{M}_n(1,\cdots,n) \nonumber\\
&+\left( \epsilon_i\epsilon_{i-1} \frac{\Delta_{i-1}-J_{i-1}+\bar{z}_{i-1,i}\bar{\partial}_{i-1}}{z_{i-1,i}}\right) e^{\frac{\partial}{\partial\Delta_i}}\mathcal{M}_n(1,\cdots,n)= 0 \, , \label{eq:nptBGshift}
\end{align}
together with the one obtained by $i-1\leftrightarrow i+1$.
The other element that we need to translate Eq. (\ref{eq:nptBGshift}) into its Carrollian version is a dictionary between $\Delta_i$ and some operator acting on the Carrollian amplitudes.
For celestial amplitudes, $\Delta_i$ corresponds to $-\omega_i\partial_{\omega_i}$ acting on the amplitudes in momentum basis \cite{Kapec:2016jld,Adamo:2019ipt,Guevara:2019ypd,Fotopoulos:2020bqj}. Therefore, translating into the operator acting on the Carrollian operator, we find
\begin{align}
&\text{\ \ Celestial}  \qquad\qquad\qquad\qquad \text{Carrollian} \nonumber\\
&\Delta_i \mathcal{M}_n \longleftrightarrow -\omega_i\partial_{\omega_i} \mathcal{A}_n \longleftrightarrow \partial_{u_i}\left( u_i C_n\right) \, . \label{eq:Delta}
\end{align} Again, this correspondence can also be directly checked using \eqref{Btransfom2}.

By using Eqs.(\ref{eq:etopartialDelta}), (\ref{eq:nptBGshift}), and (\ref{eq:Delta}), we find the differential equations satisfied by the Carrollian MHV gluon amplitudes:
\begin{align}
&\left( \partial_i -\frac{1+u_i\partial_{u_i}}{z_{i-1,i}} -\frac{1}{z_{i+1,i}}\right)\partial_{u_{i-1}}C_n(1,\cdots,n) \nonumber\\
&+ \left(  \frac{1+u_{i-1}\partial_{u_{i-1}} -J_{i-1}+\bar{z}_{i-1,i} \, \bar{\partial}_{i-1}}{z_{i-1,i}} \right) \partial_{u_{i}} C_n(1,\cdots,n) = 0 \, ,\label{eq:CarrollBG}
\end{align}
together with the one with $i-1\leftrightarrow i+1$. 

A few remarks are in order before we proceed. From Eq.(\ref{eq:CarrollBG}), we can obtain differential equations satisfied by $\overline{\text{MHV}}$ Carrollian gluon amplitudes by switching $z$ and $\bar{z}$, and flipping $J$ to $-J$,
\begin{align}
&\left( \bar{\partial}_i -\frac{1+u_i\partial_{u_i}}{\bar{z}_{i-1,i}} -\frac{1}{\bar{z}_{i+1,i}}\right)\partial_{u_{i-1}}C_{n,\overline{\text{MHV}}}(1,\cdots,n) \nonumber\\
&+ \left(  \frac{1+u_{i-1}\partial_{u_{i-1}} +J_{i-1}+{z}_{i-1,i} \, \partial_{i-1}}{\bar{z}_{i-1,i}} \right) \partial_{u_{i}} C_{n,\overline{\text{MHV}}}(1,\cdots,n) = 0 \, ,\label{eq:CarrollBGMHVbar}
\end{align}
together with the one with $i-1\leftrightarrow i+1$. 

Another notice is that we can generalize the Carrollian differential equations to arbitrary $u$-descendants. For example, if we take a derivative with respect to $u_j$ from Eq.(\ref{eq:CarrollBG}), we find
\begin{align}
&\left(\partial_i -\frac{1+\delta_{ij}+u_i\partial_{u_i}}{z_{i-1,i}} -\frac{1}{z_{i+1,i}}\right) \partial_{u_{i-1}}\partial_{u_j}C_n \nonumber\\
&+\left( \frac{1+\delta^{j}_{i-1} + u_{i-1}\partial_{u_{i-1}} -J_{i-1}+\bar{z}_{i-1,i} \, \bar{\partial}_{i-1}}{z_{i-1,i}}\right) \partial_{u_i}\partial_{u_j} C_n = 0 \, . \label{eq:CBGu1}
\end{align}
It is straightforward to generalize it to the whole tower of the $u$-descendants Eq.(\ref{eq:Cudescendants}),
\begin{align}
&\left(\partial_i -\frac{1+m_i+u_i\partial_{u_i}}{z_{i-1,i}}-\frac{1}{z_{i+1,i}}\right)\partial_{u_{i-1}}C_n^{m_1\dots m_n} \nonumber\\
&+ \left(\frac{1+m_{i-1}+u_{i-1}\partial_{u_{i-1}}-J_{i-1}+\bar{z}_{i-1,i}\bar{\partial}_{i-1}}{z_{i-1,i}} \right)\partial_{u_i}C_n^{m_1\dots m_n} = 0 \, . \label{eq:CBGu2}
\end{align}

\paragraph{Examples of solutions} There exist two different types of Carrollian CFT correlators compatible with the Ward identities: the time-dependent correlators, referred to as electric (timelike), and the time-independent correlators, called  magnetic (spacelike). The latter reduce to correlators of a standard relativistic CFT living in one dimension lower. For instance, the two-point function in three dimensions have the following two branches of solutions for the Ward identities \cite{Chen:2021xkw,deBoer:2021jej,Baiguera:2022lsw,Rivera-Betancour:2022lkc,deBoer:2023fnj}: 
\begin{equation}
\label{elec vs magn}
    \langle \Phi_{(k_1,\bar k_1)}(u_1, z_1, \bar{z}_1) \Phi_{(k_2,\bar k_2)}(u_2, z_2, \bar{z}_2) \rangle  = \left\{
    \begin{array}{ll}
        \frac{\alpha}{u_{12}^{k_1 + k_2 +\bar{k}_1 + \bar{k}_2 -2}}\delta^{(2)}(z_{12}) \delta_{k_1 + k_2,\bar{k}_1 + \bar{k}_2}  \\
        \frac{\beta}{z_{12}^{k_1+ k_2} \bar{z}_{12}^{\bar{k}_1+ \bar{k}_2}} \delta_{k_1,k_2}\delta_{\bar{k}_1,\bar{k}_2} 
    \end{array}
\right. 
\end{equation} where $\alpha$ and $\beta$ are normalization constants, and $(k_i, \bar{k}_i)$ are the Carrollian weights. The first branch is $u$-dependent and involves a $\delta$-function with respect to $z$ and $\bar{z}$. The second branch is $u$-independent and coincides with the two-point correlation function in a two-dimensional CFT. While the electric branch has been shown to be relevant to describe a scattering process involving massless finite-energy particles, the precise role of the magnetic branch remains unclear. Magnetic correlation functions seem to appear in the soft sector (see e.g. \cite{Himwich:2020rro,Pasterski:2021dqe,Freidel:2022skz,Fiorucci:2023lpb}), and this observation is buttressed by taking the inverse Fourier transform of the $u$-independent branch in \eqref{elec vs magn}:
\begin{multline}
\int^{+\infty}_{-\infty} du_1 e^{-i\omega_1 u_1} \int^{+\infty}_{-\infty} du_2 e^{-i\omega_1 u_1}  \frac{\beta}{z_{12}^{k_1+ k_2} \bar{z}_{12}^{\bar{k}_1+ \bar{k}_2}} \delta_{k_1,k_2}\delta_{\bar{k}_1,\bar{k}_2} \\
 = \delta (\omega_1 ) \delta (\omega_2 )  \frac{(2\pi)^2 \beta}{z_{12}^{k_1+ k_2} \bar{z}_{12}^{\bar{k}_1+ \bar{k}_2}} \delta_{k_1,k_2}\delta_{\bar{k}_1,\bar{k}_2}
 \label{inverse fourier}
\end{multline} (the first particle being seen as outgoing and the second as incoming). We leave a complete characterization of the magnetic branch in the context of scattering amplitudes for future work.

For higher-point cases, there will be more branches, see e.g., \cite{Nguyen:2023miw,Bagchi:2023cen}.
In any case, we find that any correlators from the magnetic branch would automatically satisfy the Carrollian differential equations (\ref{eq:CarrollBG}) since $C_n$ always appears with an operator $\partial_u$ acting on it. Here we shall give a simple example of the three-point case. One can check that
\be
C_3(-,-,+)_{\text{Mag}} = \frac{z_{12}}{z_{13}z_{23} \bar{z}_{12}^2} \,  \label{eq:C3Mag}
\ee
satisfies Eq.(\ref{eq:CarrollBG}) with the correct conformal weights of each operator as the conformal dimension $\Delta = k + \bar{k}$ of each operator has to be $1$ (see \eqref{carrollian weights}). As we will see, Eq.(\ref{eq:C3Mag}) will also appear in the non-trivial three-point solution of Eq.(\ref{eq:CarrollBG}) as a prefactor.


As another simple example, one can check that the three-point Carrollian MHV gluon amplitudes obtained from the flat-space amplitudes in $(2,2)$ signature satisfy the Carrollian BG Eq.(\ref{eq:CarrollBG}). Consider the case where particle 1 and 2 are incoming, particle 3 is outgoing, one finds \cite{Mason:2023mti}
\begin{align}
C_3(-,-,+)_{\text{flat}} &= \int_0^{\infty} d\omega_1 d\omega_2 d\omega_3 e^{-i\omega_1 u_1} e^{-i\omega_2 u_2} e^{i\omega_3 u_3} \frac{\langle 12\rangle^3}{ \langle 23\rangle \langle 31\rangle}  \delta^{(4)}(\omega_1q_1+\omega_2q_2-\omega_3q_3)\nonumber\\
&4 \int_0^{\infty} d\omega_1 d\omega_2 d\omega_3 e^{-i\omega_1 u_1} e^{-i\omega_2 u_2} e^{i\omega_3 u_3} \frac{\omega_1\omega_2}{\omega_3} \frac{z_{12}^3}{z_{23}z_{31}} \nonumber\\
&\qquad \qquad  \times \frac{1}{\omega_3^2 \, z_{23} z_{31}} \delta\left( \omega_1 -\frac{z_{32}}{z_{12}} \omega_3\right) \delta\left( \omega_2 -\frac{z_{31}}{z_{21}}\omega_3\right) \delta(\bar{z}_{13})\delta(\bar{z}_{23}) \nonumber\\
&=4\int_0^\infty d\omega_3 \exp\left[i\omega_3\left( u_3- \frac{z_{32}}{z_{12}} u_1 -\frac{z_{31}}{z_{21}} u_2 \right) \right] \frac{1}{\omega_3} \frac{z_{12}}{z_{23}z_{31}}\delta(\bar{z}_{13})\delta(\bar{z}_{23}) \nonumber\\
&= \frac{4z_{12}}{z_{23}z_{31}} \delta(\bar{z}_{13})\delta(\bar{z}_{23}) \mathcal{I}_0\left( \frac{z_{32}}{z_{12}}u_1+\frac{z_{31}}{z_{21}}u_2-u_3\right) \, , \label{eq:C3flat}
\end{align}
where $\mathcal{I}_0(x)$ is defined by
\be
\mathcal{I}_0(x) = \int_0^{+\infty}  \frac{d\omega}{\omega}e^{-i\omega x} \, , \label{eq:I0u12}
\ee
which is divergent, but can be regularized \cite{Mason:2023mti}.
\be
\mathcal{I}_\beta(x) = \lim_{\epsilon\rightarrow 0^+} \int_0^{+\infty} d\omega\, \omega^{\beta-1} e^{-i\omega x-\omega \epsilon} = \lim_{\epsilon\rightarrow 0^+} \frac{\Gamma(\beta)(-i)^{\beta}}{(x-i\epsilon)^{\beta}} \, . \label{eq:Mellin_E_to_omegax}
\ee
In the limit $\beta\rightarrow 0^+$, one finds
\be
\mathcal{I}_{\beta}(x) = \frac{1}{\beta}-\left(\gamma_E+\ln|x| +\frac{i\pi}{2}\text{sign}(x)\right) +\mathcal{O}(\beta) \, ,
\ee
where $\gamma_E$ is the Euler-Mascheroni constant. One can see that the divergent piece would not depend on the $u$ coordinates, meaning it would be automatically killed by the $\partial_u$ in (\ref{eq:CarrollBG}).
We can check that Eq.(\ref{eq:C3flat}) satisfies the differential equations Eq.(\ref{eq:CBGu1}), by choosing $i=j=3$, and using the identities such as $\bar{z}_{23} \partial_{\bar{z}_2} \delta(\bar{z}_{23}) = -\bar{z}_{23}\delta(\bar{z}_{23})/\bar{z}_{23} = -\delta(\bar{z}_{23})$.

\subsection{Three-point solution  in the presence of a dilaton source (case I)}
In \cite{Fan:2022vbz}, the authors provided a way to compute non-distributional $n$-point celestial MHV gluon amplitudes by coupling the gluons to a massless dilaton background which breaks explicitly translation invariance. The calculation boils down to the following Mellin transforms
\be
\mathcal{M}_n(-,-,+,\cdots ,+) = \left(\prod_{i=1}^n \int_0^{+\infty} d\omega_i \, \omega_i^{\Delta_i-1} \right) \frac{\langle 12\rangle^4}{\langle 12 \rangle \langle 23 \rangle \cdots \langle n1\rangle} \frac{\delta(Q^2)}{Q^2} \, ,
\ee
where the integrand contains the Parke-Taylor formula together with $\delta(Q^2)/Q^2$. $Q$ is the sum of all momenta of gluons
\be
Q = p_1+p_2-\sum_{k=3}^n p_k \, .
\ee
where we choose particles 1 and 2 to be incoming. Our idea here is to perform similar calculations in \cite{Fan:2022vbz} by using Fourier transforms with respect to the energies $\omega$s rather than by Mellin transforms. 
\be
C_n(-,-,+,\cdots,+ ) = \left(\prod_{i=1}^n\int_0^{+\infty} d\omega_i\right) e^{-i\omega_1 u_1}e^{-i\omega_2 u_2} \left(\prod_{k=3}^n e^{i\omega_k u_k}\right) \frac{\langle 12\rangle^4}{\langle 12 \rangle \langle 23 \rangle \cdots \langle n1\rangle} \frac{\delta(Q^2)}{Q^2} \, . \label{eq:CndeltaQ2}
\ee
It would give us a solution of the Carrollian BG equations (\ref{eq:CarrollBG}). We begin with the simplest case, which is the three-point:
\be
C_3(-,-,+) = \int_0^{+\infty} d\omega_1 d\omega_2 d\omega_3 e^{-i\omega_1 u_1}e^{-i\omega_2 u_2} e^{i\omega_3 u_3} \frac{\omega_1\omega_2}{\omega_3} \frac{z_{12}^3}{z_{23}z_{31}} \frac{\delta(Q^2)}{Q^2} \, , \label{eq:C3deltaQ2}
\ee
where
\be
\frac{\delta(Q^2)}{Q^2} = \frac{\delta'\left(\omega_3-\frac{\omega_1\omega_2 z_{12}\bar{z}_{12}}{\omega_1 z_{13}\bar{z}_{13}+\omega_2 z_{23}\bar{z}_{23}}\right)}{(\omega_1 z_{13}\bar{z}_{13} +\omega_2 z_{23} \bar{z}_{23})^2} . 
\ee
See section 3 in \cite{Fan:2022vbz} for details. Here $\delta'(x)$ is the derivative of the $\delta$ function.

Actually it turns out that it is simpler to compute $\partial_{u_3} C_3$. Once we have $\partial_{u_3} C_3$, we can integrate it over $u_3$ to get $C_3$ modulo a constant of the integration that we will neglect. Starting from Eq.(\ref{eq:C3deltaQ2}), we compute
\begin{align}
\partial_{u_3} C_3(-,-,+) = i \int_0^{+\infty} d\omega_1 d\omega_2 d\omega_3 e^{-i\omega_1 u_1}e^{-i\omega_2 u_2} e^{i\omega_3 u_3} \omega_1\omega_2  \frac{z_{12}^3}{z_{23}z_{31}}  \frac{\delta'\left(\omega_3-\frac{\omega_1\omega_2 z_{12}\bar{z}_{12}}{\omega_1 z_{13}\bar{z}_{13}+\omega_2 z_{23}\bar{z}_{23}}\right)}{(\omega_1 z_{13}\bar{z}_{13} +\omega_2 z_{23} \bar{z}_{23})^2} \, .
\end{align}
Integrating it by part over $\omega_3$ and using the $\delta$-function to fix $\omega_3$, we find
\begin{align}
\partial_{u_3} C_3(-,-,+)  = -\frac{z_{12}^3}{z_{23}z_{13}} u_3 \int_0^{+\infty} d\omega_1 d\omega_2 & e^{-i\omega_1 u_1}e^{-i\omega_2 u_2} \exp\left( i u_3 \frac{\omega_1\omega_2 z_{12} \bar{z}_{12}}{\omega_1 z_{13} \bar{z}_{13} +\omega_2 z_{23} \bar{z}_{23}}\right)  \nonumber\\
&\times\frac{\omega_1\omega_2}{(\omega_1 z_{13}\bar{z}_{13} +\omega_2 z_{23} \bar{z}_{23})^2} \, . \label{eq:pu3C3}
\end{align}
To evaluate the $d\omega_1$ and $d\omega_2$ integrals, first we expand the exponential with $u_3$:
\begin{align}
&\partial_{u_3} C_3(-,-,+) \nonumber\\
 =& -\frac{z_{12}^3}{z_{23}z_{13}} u_3 \int_0^{+\infty} d\omega_1 d\omega_2  e^{-i\omega_1 u_1}e^{-i\omega_2 u_2}\frac{\omega_1\omega_2}{(\omega_1 z_{13}\bar{z}_{13} +\omega_2 z_{23} \bar{z}_{23})^2} \nonumber\\
&\qquad \qquad\times \sum_{n=0}^{\infty} \frac{\left( i u_3 \frac{\omega_1\omega_2 z_{12} \bar{z}_{12}}{\omega_1 z_{13} \bar{z}_{13} +\omega_2 z_{23} \bar{z}_{23}}\right)^n}{n!} \nonumber\\
=& -\frac{z_{12}^3}{z_{23}z_{13}} \sum_{n=0}^{\infty} \frac{u_3 (iu_3 z_{12} \bar{z}_{12})^n}{n!} \int_0^{+\infty} d\omega_1 d\omega_2 e^{-i\omega_1 u_1}e^{-i\omega_2 u_2} \frac{\omega_1^{n+1}\omega_2^{n+1}}{(\omega_1 z_{13}\bar{z}_{13} +\omega_2 z_{23} \bar{z}_{23})^{n+2}} \nonumber\\
=&-\frac{z_{12}^3}{z_{23}z_{13}} \sum_{n=0}^{\infty} \frac{u_3 (iu_3 z_{12} \bar{z}_{12})^n}{n!} I_{n3} \, , \label{eq:pu3C3step2}
\end{align}
where we denote the integrals of $\omega_1$ and $\omega_2$ as $I_{n3}$. Next, we make the following change of variables
\be
\omega_P= \omega_1+\omega_2 \, , \quad t= \frac{\omega_1}{\omega_P} \, . \label{eq:w1w2totwp}
\ee
$I_{n3}$ becomes
\begin{align}
I_{n3} =& \int_0^1 dt \int_0^{+\infty} d\omega_P \frac{\omega_P \omega_P^{n+1} t^{n+1} \omega_P^{n+1} (1-t)^{n+1}}{(\omega_P t z_{13} \bar{z}_{13} +\omega_P (1-t) z_{23}\bar{z}_{23})^{n+2}} e^{-i\omega_P( t u_1 +(1-t) u_2)} \nonumber\\
=& \int_0^{1}dt \frac{ t^{n+1} (1-t)^{n+1}}{(t z_{13}\bar{z}_{13} +(1-t) z_{23}\bar{z}_{23})^{n+2}} \int_0^{+\infty} d\omega_P \,\omega_P^{n+1} e^{-i\omega_P( t u_1 +(1-t) u_2)}\nonumber\\
=& \int_0^{1}dt \frac{ t^{n+1} (1-t)^{n+1}}{(t z_{13}\bar{z}_{13} +(1-t) z_{23}\bar{z}_{23})^{n+2}} \frac{\Gamma(2+n)}{(i t u_1+i(1-t) u_2)^{n+2}} \, . \label{eq:In3}
\end{align}
To evaluate the $dt$ integral of $I_{n3}$, we perform the change of variable:
\be
\tilde{t} = \frac{t}{t-1} \, . \label{eq:tott}
\ee
Eq.(\ref{eq:In3}) becomes
\begin{align}
I_{n3} &= -\Gamma(2+n) \int_0^{-\infty} \frac{d\tilde{t}}{(1-\tilde{t}\, )^2} \left( \frac{-\tilde{t}}{1-\tilde{t}}\right)^{n+1} \left( \frac{1}{1-\tilde{t}} \right)^{n+1} \nonumber\\
&\quad \quad \times \left( \frac{-\tilde{t}}{1-\tilde{t}} |z_{13}|^2 +\frac{1}{1-\tilde{t}} |z_{23}|^2\right)^{-n-2} \left(i \frac{-\tilde{t}}{1-\tilde{t}} \,u_1 +i \frac{1}{1-\tilde{t}} \, u_2\right)^{-n-2}  \nonumber\\
&=\Gamma(2+n) i^{-n-2} \int_0^{+\infty} dt \, t^{n+1}\, ( |z_{23}|^2 +t |z_{13}|^2)^{-n-2} (u_2 +t u_1)^{-n-2} \, ,
\end{align}
where in the last line we rename $-\tilde{t}$ as $t$. At this stage, the last line takes the form as an integral representation of the hypergeometric function $_2F_1$. We find
\be
I_{n3} = i^{-n-2} u_2^{-2-n} (z_{13}\bar{z}_{13})^{-2-n} \frac{\Gamma^3(2+n)}{\Gamma(4+2n)} \, _2F_1\left(2+n,2+n,4+2n;1-\frac{u_1|z_{23}|^2}{u_2 |z_{13}|^2}\right) \, .
\ee
Plug it back to Eq.(\ref{eq:pu3C3step2}), we have
\begin{align}
\partial_{u_3}C_3(-,-,+) =& -\frac{z_{12}^3}{z_{23}z_{13}}\sum_{n=0}^\infty \frac{u_3^{n+1}(i z_{12}\bar{z}_{12})^n}{n!} \frac{1}{(i u_2 z_{13}\bar{z}_{13})^{n+2}} \nonumber\\
& \qquad\qquad\qquad \times\frac{\Gamma^3(2+n)}{\Gamma(4+2n)} \, _2F_1\left(2+n,2+n,4+2n;1-\frac{u_1|z_{23}|^2}{u_2 |z_{13}|^2}\right)
\end{align}
Integrating it over $u_3$, we obtain
\begin{align}
&C_3(-,-,+) \nonumber\\
=& \frac{z_{12}}{z_{13}z_{23}\bar{z}^2_{12}} \sum_{n=0}^\infty \left(\frac{u_3 |z_{12}|^2}{u_2 |z_{13}|^2} \right)^{n+2} \frac{(n+1)\Gamma^2(2+n)}{(n+2)\Gamma(4+2n)}\, _2F_1\left(2+n,2+n,4+2n; 1-\frac{u_1|z_{23}|^2}{u_2 |z_{13}|^2}\right) \, . \label{eq:C3result1}
\end{align}
As we mentioned before, the prefactor is given by Eq.(\ref{eq:C3Mag}). 
Moreover, we check numerically that the sum is convergent in the region where $\left|1-\frac{u_1|z_{23}|^2}{u_2 |z_{13}|^2}\right|<1$.

Using the following identity
\begin{align}
&\, _2F_1(n+1,n+m+1,n+m+l+2;z) \nonumber\\
=& \frac{(n+m+l+1)!(-1)^m}{l!\,n!\,(n+m)!(m+l)!} \frac{d^{n+m}}{dz^{n+m}}\left( (1-z)^{m+l}\frac{d^l}{dz^l} \, _2F_1(1,1,2,z)\right) \, \label{Erdelyi}
\end{align}
where 
\be\label{HyperB}
\,_2F_1(1,1,2,z) = -\frac{\log(1-z)}{z} \, ,
\ee
we can rewrite Eq.(\ref{eq:C3result1}) as 
\begin{align}
&C_3(-,-,+) \nonumber\\
=& -\frac{z_{12}}{z_{13}z_{23}\bar{z}^2_{12}} \sum_{n=0}^\infty \left(\frac{u_3 |z_{12}|^2}{u_2 |z_{13}|^2} \right)^{n+2} \frac{1}{n!(n+2)!}\; \frac{d^{n+1}}{dy^{n+1}}\left\{ (1-y)^{n+1}\frac{d^{n+1}}{dy^{n+1}}\frac{\log(1-y)}{y}\right\} \, , \label{eq:C3result1diff}
\end{align}
where we define
\be\label{Variable}
y= 1-\frac{u_1|z_{23}|^2}{u_2 |z_{13}|^2} \, .
\ee

Notice that the sum of Eq.(\ref{eq:C3result1}) is written in a way that the symmetric property under $1\leftrightarrow2$ is not manifest. To see indeed the sum is symmetric under $1\leftrightarrow2$, we use the following identity of the hypergeometric function $_2F_1$,
\be
\, _2F_1(\alpha,\beta,2\beta;z) = \left( 1-\frac{z}{2}\right)^{-\alpha} \, _2F_1\left( \frac{\alpha}{2}, \frac{\alpha+1}{2}, \beta +\frac{1}{2}; \left( \frac{z}{2-z}\right)^2 \right) \,  \label{eq:2F1id1}
\ee
to rewrite Eq.(\ref{eq:C3result1}) as 
\begin{align}
&C_3(-,-,+) \nonumber\\
=&\frac{z_{12}}{z_{13}z_{23}\bar{z}^2_{12}} \sum_{n=0}^\infty \left( \frac{2 u_3 |z_{12}|^2}{u_2 |z_{13}|^2 +u_1 |z_{23}|^2}\right)^{n+2} \frac{(n+1)\Gamma^2(2+n)}{(n+2)\Gamma(4+2n)}\nonumber\\
& \qquad\qquad\qquad \times \, _2F_1\left( \frac{n+2}{2}, \frac{n+3}{2}, n+\frac{5}{2}; \left( \frac{u_2|z_{13}|^2 -u_1 |z_{23}|^2}{u_2|z_{13}|^2+u_1|z_{23}|^2}\right)^2\right) \, , \label{eq:C3result2}
\end{align}
where the sum is manifestly symmetric under $1\leftrightarrow 2$ as expected.

{\it Special case 1:} We consider the case where $ z_{13}\bar{z}_{13} = a u_1$ and $z_{23}\bar{z}_{23} = a u_2$, with $a$ being a constant. Then $I_{n3}$ Eq.(\ref{eq:In3}) becomes
\begin{align}
I_{n3}(\text{case 1}) &= \frac{1}{i^{n+2} a^{n+2}}\int_0^1 dt\, t^{n+1} (1-t)^{n+1} \frac{\Gamma(2+n)}{(t u_1 +(1-t) u_2)^{2n+4}} \nonumber\\
&=\frac{2^{-3-2n} \sqrt{\pi}}{i^{n+2} a^{n+2}} (u_1u_2)^{-n-2} \frac{\Gamma(2+n)^2}{\Gamma(\frac{5}{2}+n)} \, .
\end{align}
Repeating the same steps as we had above, we find that the three-point Carrollian amplitude in this case is given by
\begin{align}
C_3(\text{case 1}) &= \frac{z_{12}}{z_{13}z_{23}} \frac{2\sqrt{\pi}}{\bar{z}^2_{12}} \sum_{n=0}^{\infty} \left(\frac{u_3 z_{12}\bar{z}_{12}}{4a u_1u_2} \right)^{n+2} \frac{(n+1)\Gamma(2+n)}{(n+2)\Gamma\left(\frac{5}{2}+n\right)} \nonumber\\
&= \frac{4 z_{12}}{z_{13}z_{23}\bar{z}_{12}^2}\frac{\sin^{-1}\left(\sqrt{\frac{u_3 z_{12}\bar{z}_{12}}{4a u_1u_2}} \right)}{\sqrt{1-\frac{u_3 z_{12}\bar{z}_{12}}{4a u_1u_2}}}\left(\sqrt{\frac{u_3 z_{12}\bar{z}_{12}}{4a u_1u_2}}- \sqrt{1-\frac{u_3 z_{12}\bar{z}_{12}}{4a u_1u_2}}\sin^{-1}\left( \sqrt{\frac{u_3 z_{12}\bar{z}_{12}}{4a u_1u_2}}\right)\right) \, .
\end{align}
This special case sugguests that the generic formula for $C_3$ and its $u$-descendants  might be resummed into a compact formula as well.

It turns out that we can start from Eq.(\ref{eq:pu3C3}),  and perform change of variables as we had in (\ref{eq:w1w2totwp}). Eq.(\ref{eq:pu3C3}) becomes
\begin{align}
\partial_{u_3} C_3(-,-,+)  &= -\frac{z_{12}^3}{z_{23}z_{13}} u_3 \int_0^{1}dt \int_0^{+\infty} d\omega_P\, \omega_P\, e^{-i t\omega_P u_1} e^{-i(1-t) \omega_P u_2} \frac{t(1-t)}{(tz_{13}\bar{z}_{13} +(1-t) z_{23}\bar{z}_{23})^2} \nonumber\\
&=\exp\left(i u_3 \frac{t(1-t) \omega_P z_{12}\bar{z}_{12}}{(t z_{13}\bar{z}_{13} +(1-t) z_{23}\bar{z}_{23}}\right) \nonumber\\
&=\frac{z_{12}^3}{z_{23}z_{13}} u_3 \int_0^1 dt \frac{t(1-t)}{(tz_{13}\bar{z}_{13} +(1-t) z_{23}\bar{z}_{23})^2} \frac{1}{\left( u_3\frac{t(1-t)  z_{12}\bar{z}_{12}}{t z_{13}\bar{z}_{13} +(1-t) z_{23}\bar{z}_{23}} - t u_1-(1-t) u_2\right)^2}  \nonumber\\
&=\frac{z_{12}^3}{z_{23}z_{13}} u_3 \int_0^1 dt \frac{t(1-t)}{\left( u_3 t(1-t) z_{12}\bar{z}_{12}-(t u_1 +(1-t)u_2)(t z_{13}\bar{z}_{13} +(1-t) z_{23}\bar{z}_{23})\right)^2}
\end{align}
We perform the change of variale as we had in Eq.(\ref{eq:tott}), it becomes
\begin{align}
\partial_{u_3} C_3(-,-,+)  &=\frac{z_{12}^3}{z_{23}z_{13}} u_3 \int_0^{-\infty} d\tilde{t} \frac{\tilde{t}}{\left( -\tilde{t}\,u_3  z_{12}\bar{z}_{12}-(-\tilde{t}\, u_1 +u_2)(-\tilde{t}\, z_{13}\bar{z}_{13} + z_{23}\bar{z}_{23})\right)^2} \nonumber\\
&=\frac{z_{12}^3}{z_{23}z_{13}} u_3 \int_0^{+\infty} dt \frac{t}{\left[ t u_3 z_{12}\bar{z}_{12} -(t u_1 +u_2) (t z_{13}\bar{z}_{13} +z_{23}\bar{z}_{23})\right]^2} \, ,
\end{align}
where we rename $-\tilde{t}$ as $t$ in the last line. The intgral takes the form \cite{Gradshteyn:1943cpj},
\begin{align}
&\int_0^{+\infty} \frac{tdt}{(a t^2 +2 b t+ c)^2}  \nonumber\\
=& \frac{1}{2(ac-b^2)} -\frac{b}{2(ac-b^2)^{\frac{3}{2}}} \,\text{arccot} \frac{b}{\sqrt{ac-b^2}} \quad \text{for} \, ac>b^2 \nonumber\\ 
=&\frac{1}{2(ac-b^2)} + \frac{b}{4(b^2-ac)^{\frac{3}{2}}} \ln \frac{b+\sqrt{b^2-ac}}{b-\sqrt{b^2-ac}} \quad \text{for} \, b^2>ac>0 \nonumber\\
=& \frac{1}{6 b^2}\qquad \qquad \qquad \qquad \qquad \qquad \qquad \qquad \quad \, \, \text{for} \, ac=b^2 \, ,
\end{align}
As a result, we find
\begin{align}
\partial_{u_3} C_3(-,-,+)&=\frac{z_{12}^3}{z_{23}z_{13}} u_3 \left( \frac{1}{2(ac-b^2)} -\frac{b}{2(ac-b^2)^{\frac{3}{2}}} \,\text{arccot} \frac{b}{\sqrt{ac-b^2}}\right) \quad \text{for} \, ac>b^2 \nonumber\\ 
&=\frac{z_{12}^3}{z_{23}z_{13}} u_3\left(\frac{1}{2(ac-b^2)} + \frac{b}{4(b^2-ac)^{\frac{3}{2}}} \ln \frac{b+\sqrt{b^2-ac}}{b-\sqrt{b^2-ac}}\right) \quad \text{for} \, b^2>ac>0 \nonumber\\ 
&=\frac{z_{12}^3}{z_{23}z_{13}} u_3\frac{1}{6 b^2}\qquad \qquad \qquad \qquad \qquad \qquad \qquad \qquad \quad \, \, \text{for} \, ac=b^2 \, , \label{eq:pu3C3Grad}
\end{align}
where
\begin{align}
a&=u_1 z_{13}\bar{z}_{13} \, ,  \\
b&= \frac{u_1 z_{23}\bar{z}_{23} +u_2 z_{13}\bar{z}_{13} -u_3 z_{12}\bar{z}_{12}}{2} \, ,\\
c&= u_2 z_{23}\bar{z}_{23} \, .
\end{align}
The result given by Eq.(\ref{eq:pu3C3Grad}) is anti-symmetric under exchanging $1\leftrightarrow2$ as expected.
One can try to integrate it over $u_3$ to get a formula for $C_3(-,-,+)$.
On the other hand, we have differential equations satisfied by $\partial_{u_3}C_3(-,-,+)$ given by Eqs.(\ref{eq:CBGu1}) and (\ref{eq:CBGu2}). Having a close formula for $\partial_{u_3}C_3(-,-,+)$ is equally good.

\subsection{Three-point solution  in the presence of a dilaton source (case II)}
We can also use the massless dilaton source in \cite{Stieberger:2022zyk}. For the three-point case, we shall compute
\begin{align}
C_{3L} &= \int_0^{\infty} d\omega_1 d\omega_2 d\omega_3 e^{i u_1\omega_1}e^{i u_2 \omega_2} e^{i u_3 \omega_3} \frac{\langle 12\rangle^3}{\langle 23\rangle \langle 31\rangle} \frac{1}{Q^2} \nonumber\\
&= \frac{z_{12}^3}{z_{23}z_{31}} \int_0^{\infty} d\omega_1 d\omega_2 d\omega_3 e^{i u_1\omega_1}e^{i u_2 \omega_2} e^{i u_3 \omega_3} \frac{\omega_1\omega_2}{\omega_3} \frac{1}{\omega_1\omega_2 z_{12}\bar{z}_{12} +\omega_2\omega_3 z_{23}\bar{z}_{23} +\omega_1\omega_3 z_{13}\bar{z}_{13}} \, ,
\end{align}
where particles 1, 2, and 3 are all outgoing.
Similar to the calculations that we did before, it is easier to compute the $u-$ descendants,
\begin{align}
\partial_{u_3} C_{3L} &= \frac{z_{12}^3}{z_{23}z_{31}} i\int_0^{\infty} d\omega_1 d\omega_2 d\omega_3 e^{i u_1\omega_1 +i u_2 \omega_2 +i u_3 \omega_3} \frac{\omega_1\omega_2}{\omega_1\omega_2 z_{12}\bar{z}_{12} +\omega_2\omega_3 z_{23}\bar{z}_{23} +\omega_1\omega_3 z_{13}\bar{z}_{13}} \nonumber\\
&=\frac{z_{12}^3}{z_{23}z_{31}} i\int_0^\infty d\omega_1 d\omega_2 e^{i u_1 \omega_1 +i u_2\omega_2} \frac{\omega_1\omega_2}{\omega_1 z_{13}\bar{z}_{13} +\omega_2 z_{23}\bar{z}_{23}} \, \exp\left( -i u_3 \frac{\omega_1\omega_2 z_{12}\bar{z}_{12}}{\omega_1 z_{13}\bar{z}_{13} +\omega_2 z_{23}\bar{z}_{23}}\right) \nonumber\\
&\quad\qquad\qquad\qquad \times\Gamma\left(0,-i u_3\frac{ \omega_1\omega_2 z_{12}\bar{z}_{12}}{\omega_1 z_{13} \bar{z}_{13} +\omega_2 z_{23}\bar{z}_{23}} \right) \, , \label{eq:puC3L}
\end{align}
where $\Gamma(\alpha,x)$ is the incomplete gamma function defined as
\be
\Gamma(\alpha,x) = \int_x^\infty dt\, e^{-t}\, t^{\alpha-1} \, .
\ee
Next, we make a change of variable as we had in Eq.(\ref{eq:w1w2totwp}), then Eq.(\ref{eq:puC3L}) becomes
\begin{align}
\partial_{u_3} C_{3L} &=  \frac{z_{12}^3}{z_{23}z_{31}} i \int_0^1 dt \, \int_0^{+\infty} d\omega_P \, \omega_P^2 \frac{t(1-t)}{t z_{13}\bar{z}_{13} +(1-t) z_{23}\bar{z}_{23}} e^{i\omega_P t\, u_1} e^{i\omega_P (1-t) u_2} \nonumber\\
&\qquad\qquad \times \exp\left(-i u_3\, \frac{\omega_P z_{12}\bar{z}_{12} \, t(1-t)}{t z_{13}\bar{z}_{13} +(1-t) z_{23}\bar{z}_{23}} \right) \Gamma\left(0,\, -i u_3\, \frac{\omega_P z_{12}\bar{z}_{12} \, t(1-t)}{t z_{13}\bar{z}_{13} +(1-t) z_{23}\bar{z}_{23}} \right) \, ,
\end{align}
where the $\omega_P$ integral can be evaluated by using Eq.(6.455) of Gradshteyn and Ryzhik 7th edition,
\be
\int_0^{+\infty} dx x^{\mu-1} e^{-\beta x} \,\Gamma(\nu, \alpha x) = \frac{\alpha^{\nu} \Gamma(\mu+\nu)}{\mu (\alpha+\beta)^{\mu+\nu}} \, _2F_1\left( 1, \mu+\nu,\mu+1;\frac{\beta}{\alpha+\beta}\right) \, .
\ee
We find
\begin{align}
\partial_{u_3}C_{3L} &=  \frac{z_{12}^3}{z_{23}z_{31}} \frac{2}{3} \int_0^1 dt \, \frac{t(1-t)}{t z_{13}\bar{z}_{13} +(1-t) z_{23}\bar{z}_{23}} \frac{1}{(t u_1 +(1-t) u_2)^3} \nonumber\\
&\qquad \qquad \times \, _2F_1\left( 1,3,4; \, 1-\frac{u_3 t(1-t) z_{12}\bar{z}_{12}}{(tu_1 +(1-t) u_2)(t z_{13} \bar{z}_{13} +(1-t) z_{23}\bar{z}_{23})}\right) \, .
\end{align}
We make a change of variable as we had in Eq.(\ref{eq:tott}), it becomes
\begin{align}
\partial_{u_3}C_{3L} &= \frac{2}{3} \frac{z_{12}^3}{z_{23}z_{31}} \int_0^\infty dt \frac{t}{(t z_{13}\bar{z}_{13} +z_{23}\bar{z}_{23})(t u_1+u_2)^3} \nonumber\\
&\qquad \qquad \times \, _2F_1\left( 1,3,4; \, 1- \frac{t u_3 z_{12}\bar{z}_{12}}{(t u_1+u_2)(t z_{13}\bar{z}_{13} +z_{23}\bar{z}_{23})} \right) \, . \label{eq:puC3Lstep2}
\end{align}
The hypergeometric function in Eq.(\ref{eq:puC3Lstep2}) admits the following series expansion,
\be
\, _2F_1(1,3,4;x) = 3 \sum_{n=0}^\infty \frac{x^n}{n+3}=3\ \Phi(x,1,3) \, ,
\ee
which in turn is related to Lerch function $\Phi$.
Therefore, Eq.(\ref{eq:puC3Lstep2}) can be written as 
\begin{align}
\partial_{u_3}C_{3L} &= \frac{2\, z_{12}^3}{z_{23}z_{31}} \int_0^{+\infty} dt  \frac{t}{(t z_{13}\bar{z}_{13} +z_{23}\bar{z}_{23})(t u_1+u_2)^3} \sum_{n=0}^{\infty} \frac{\left( 1- \frac{t u_3 z_{12} \bar{z}_{12}}{(t u_1+u_2)(t z_{13}\bar{z}_{13} +z_{23} \bar{z}_{23})} \right)^n}{n+3} \, ,
\end{align}
Using the binomial expansion
\be
\left( 1- \frac{t u_3 z_{12} \bar{z}_{12}}{(t u_1+u_2)(t z_{13}\bar{z}_{13} +z_{23} \bar{z}_{23})} \right)^n = \sum_{k=0}^n (-1)^k \frac{n!}{k!(n-k)!} \frac{t^k (u_3 z_{12}\bar{z}_{12})^k}{(t u_1+u_2)^k (t z_{13} \bar{z}_{13} +z_{23}\bar{z}_{23})^k} \, ,
\ee
we obtain
\begin{align}
\partial_{u_3}C_{3L} &=\frac{2\, z_{12}^3}{z_{23}z_{31}} \sum_{n=0}^\infty \sum_{k=0}^n \int_0^{+\infty} dt \frac{1}{n+3} \, \frac{(-1)^k n!}{k! (n-k)!} \frac{t^{k+1} (u_3 z_{12} \bar{z}_{12})^k}{(t u_1+u_2)^{k+1} (t z_{13} \bar{z}_{13} + z_{23} \bar{z}_{23})^{k+3}} \nonumber\\
&=\frac{2\, z_{12}^3}{z_{23}z_{31}} \sum_{n=0}^\infty \sum_{k=0}^n  \frac{1}{n+3} \, \frac{(-1)^k n!}{k! (n-k)!} \frac{(u_3 z_{12}\bar{z}_{12})^k}{u_1 u_2} \frac{1}{(u_2 z_{13} \bar{z}_{13})^{k+1}} \frac{\Gamma^2(2+k)}{\Gamma(4+2k)}  \nonumber\\
&\qquad \qquad \qquad \times \, _2F_1\left(1+k,2+k,4+2k; \, 1-\frac{u_1 z_{23} \bar{z}_{23}}{u_2 z_{13} \bar{z}_{13}} \right) \, . \label{eq:puC3Lresult1}
\end{align}
Similar as \req{eq:C3result1diff} with \req{Erdelyi} we may cast  \req{eq:puC3Lresult1} into:
 \begin{align}
\partial_{u_3}C_{3L} &=-\frac{2\, z_{12}^3}{z_{23}z_{31}} \sum_{n=0}^\infty \sum_{k=0}^n  \frac{1}{n+3} \, \frac{(-1)^k n!}{k!(n-k)!} \frac{(u_3 z_{12}\bar{z}_{12})^k}{u_1 u_2} \frac{1}{(u_2 z_{13} \bar{z}_{13})^{k+1}}   \nonumber\\
&\times\fc{1}{k! (k+2)!}\ \fc{d^{k+1}}{dy^{k+1}} \lf\{(1-y)^{k+2}\fc{d^{k+1}}{dy^{k+1}}\, \frac{\log(1-y)}{y}\ri\}\ ,
\end{align}
with the variable \req{Variable}.
Notice that similar to Eq.(\ref{eq:C3result1}), the result \req{eq:puC3Lresult1} can be rewritten in a way that it is anti-symmetric under $1\leftrightarrow2$. Using the identity Eq.(\ref{eq:2F1id1}), $\partial_{u_3}C_{3L}$ becomes,
\begin{align}
\partial_{u_3}C_{3L} =& \frac{2 z_{12}^2}{z_{23}z_{31} \bar{z}_{12} u_1 u_2 u_3} \sum_{n=0}^\infty \sum_{k=0}^n  \frac{1}{n+3} \frac{(-1)^k n!}{k! (n-k)!} \left(\frac{2 u_3 z_{12} \bar{z}_{12}}{u_2 |z_{13}|^2 +u_1|z_{13}|^2} \right)^{k+1} \nonumber\\
& \times \frac{\Gamma^2(2+k)}{\Gamma(4+2k)} \, _2F_1\left( \frac{k+1}{2}, \frac{k+2}{2}, k +\frac{5}{2}; \, \left( \frac{u_2|z_{13}|^2 -u_1 |z_{23}|^2}{u_2|z_{13}|^2+u_1|z_{23}|^2}\right)^2\right)  \, , \label{eq:puC3Lresult2}
\end{align}
which is anti-symmetric under exchanging 1 and 2 as expected.

\subsection{Higher-point solutions}
In the presence of a generic massless dilaton source, it was shown that the higher-point non-distributional solutions of celestial BG equation can be obtained as follows \cite{Fan:2022vbz},
\be {\cal M}_N(z_1,\bz_1,\dots,z_N,\bz_N|\Delta_1,\dots,\Delta_N) ={\cal J}_N(z_i)\,{\cal S}_N(z_i,\bz_i)\label{mhv1}\ee
where
\be {\cal J}_N(z_i)=\sum_{\pi\in S_{N-2}}f^{a_1a_{\pi(2)}x_1}f^{x_1a_{\pi(3)}x_2}\cdots f^{x_{N{-}3}a_{\pi(N{-}1)}a_{N}}\frac{z_{12}^4}{z_{1\pi(2)}z_{\pi(2)\pi(3)}\cdots z_{N1}},\label{cfactor} \ee
is the holomorphic ``soft'' factor, with $(a_1,a_2,\dots,a_N)$ labeling the gluon group indices.
The Mellin transforms are contained in the ``scalar'' part
\be {\cal S}_N(z_i,\bz_i)= \label{spart}
\int {d^4X}\int_{\omega_i\geq 0}J_{\Phi}(X)\frac{e^{iX\cdot Q}}{Q^2}   d\omega_1\,  \omega_1^{\Delta_1}d\omega_2 \,\omega_2^{\Delta_2}\prod_{k\geq 3}^Nd\omega_k\,\omega_k^{\Delta_k-2},\ee
where $Q$ is the total momentum of the gluon system,
\be Q=\sum_{i=1}^NP_i \, .\ee
One can easily translate Eq.(\ref{mhv1}) to its Carrollian version,
\be
C_N(z_1,\bar{z}_1,u_1,\cdots,z_N,\bar{z}_N) = \mathcal{J}_N(z_i)\,\mathcal{CS}_N(z_i,\bar{z}_i,u_i) \, ,
\ee
where $\mathcal{J}_N(z_i)$ is given by Eq.(\ref{cfactor}), and $\mathcal{CS}_N(z_i,\bar{z}_i,u_i)$ contains all the Fourier transforms,
\be
\mathcal{CS}_N(z_i,\bar{z}_i,u_i) = \int d^4X\int_{\omega_i\geq0} J_{\Phi}(X) \frac{e^{iX\cdot Q}}{Q^2} d\omega_1  d\omega_2 \,\omega_1 \omega_2 \, e^{i\epsilon_1 \omega_1 u_1}e^{i\epsilon_2 \omega_2 u_2} \prod_{k \geq 3}^N d\omega_k \frac{1}{\omega_k} e^{i\epsilon_k \omega_k u_k} \, .
\ee
The choice of the dilaton source $J_{\Phi}(X)$ determines the integrand. For example, the dilaton source used Ref.\cite{Fan:2022vbz} is given by
\be
J_{\Phi}(X) \sim \int d^4 P\, \delta(P^2) \, e^{iP\cdot X} \,  \label{eq:JElement}
\ee
while the dilaton source used in Ref.\cite{Stieberger:2022zyk} is 
\be
J_{\Phi}(X) \sim \delta^4(X) \, . \label{eq:JLiouville}
\ee
In \cite{Stieberger:2022zyk}, it was shown that the dilaton source Eq.(\ref{eq:JLiouville}) led to celestial MHV gluon amplitudes that are related to correlators of light operators in Liouville CFT in the large central charge limit.
See also \cite{Taylor:2023bzj,Stieberger:2023fju,Melton:2023lnz,Giribet:2024vnk,Melton:2024gyu} for further developments.
Notice that in \cite{Melton:2024gyu,Melton:2023bjw}, the authors showed that the Liouville correlators appearing in the large central charge limit can be combined into the orginal celestial MHV gluon amplitudes with translation invariance. It would be interesting to see if this idea has a Carrollian description.

\section{Differential equations of Carrollian MHV graviton amplitudes} \label{sec4}
\subsection{From leading and subleading soft graviton symmetries}
\label{sec:From leading and subleading soft graviton symmetries}

In section 12 of \cite{Banerjee:2020zlg}, the authors showed two differential equations satisfied by celestial MHV graviton amplitudes derived from the null states conditions constructed from leading and subleading soft graviton symmetries,
\begin{align}
\Phi^{\sigma} &= \left[ J^{1}_{-1} P_{-1,-1}-(2\bar{h}+1)P_{-2,0}\right] G^{\sigma}_{\Delta}(z,\bar{z}) \, , \label{eq:granull1}\\
\Psi &= \left(L_{-1} P_{-1,-1} + 2 J^0_{-1} P_{-1,-1} -(\Delta+1) P_{-2,-1} -\bar{L}_{-1}P_{-2,0} \right) G^{+}_{\Delta}(z,\bar{z}) \, , \label{eq:granull2}
\end{align}
with helicity $\sigma$ and cf. \cite{Banerjee:2020zlg} for details. Inserting them into celestial MHV graviton amplitudes, one finds the following differential equations,
\begin{align}
&-\sum_{i=1}^{n-1}  \frac{(\Delta_i-J_i)(\bar{z}_i-\bar{z}_n) + (\bar{z}_i-\bar{z}_n)^2\bar{\partial}_i}{z_i-z_n} \, e^{\frac{\partial}{\partial\Delta_n}} \mathcal{M}_n(1,\cdots,n) \nonumber\\
&+(\Delta_n-J_n+1) \sum_{i=1}^{n-1} \frac{\bar{z}_i-\bar{z}_n}{z_i-z_n} \,\epsilon_i \, e^{\frac{\partial}{\partial\Delta_i}}\mathcal{M}_n(1,\cdots,n) = 0 \, , \label{eq:Mgradiff1}
\end{align}
\begin{align}
&\left( \partial_i e^{\frac{\partial}{\partial \Delta_n}} - \sum_{i=1}^{n-1} \frac{(\Delta_i-J_i)+2(\bar{z}_i-\bar{z}_n)\bar{\partial}_i}{z_i-z_n} e^{\frac{\partial}{\partial\Delta_n}}\right)\mathcal{M}_n(1,\cdots,n) \nonumber\\
&+\left( (\Delta_n+1)\sum_{i=1}^{n-1} \frac{1}{z_i-z_n} \, \epsilon_i \,e^{\frac{\partial}{\partial\Delta_i}}  + \bar{\partial}_i \sum_{i=1}^{n-1} \frac{\bar{z}_i-\bar{z}_n}{z_i-z_n} \, \epsilon_i \, e^{\frac{\partial}{\partial\Delta_i}}\right)\mathcal{M}_n(1,\cdots,n)=0 \, , \label{eq:Mgradiff2}
\end{align}
where we have chosen the last graviton $n$ to be outgoing and the helicity of it  to be $+2$.
By using Eqs.(\ref{eq:etopartialDelta}) and (\ref{eq:Delta}), we can translate Eqs.(\ref{eq:Mgradiff1}) and (\ref{eq:Mgradiff2}) into differential equations of the corresponding Carrollian graviton amplitudes,
\begin{align}
&-\sum_{i=1}^{n-1} \frac{(1+u_i\partial_{u_i}-J_i) (\bar{z}_i-\bar{z}_n) +(\bar{z}_i-\bar{z}_n)^2\bar{\partial}_i}{z_i-z_n} \partial_{u_n} C_n(1,\cdots,n)  \nonumber\\
&+(2+u_n\partial_{u_n} -J_n) \sum_{i=1}^{n-1} \frac{\bar{z}_i-\bar{z}_n}{z_i-z_n} \partial_{u_i} C_n(1,\cdots,n) =0 \, , \label{eq:Cgradiff1}
\end{align}
\begin{align}
&\left( \partial_n \partial_{u_n} - \sum_{i=1}^{n-1} \frac{(1+u_i\partial_{u_i} -J_i) +2(\bar{z}_i-\bar{z}_n) \bar{\partial}_i}{z_i-z_n} \, \partial_{u_n}\right)C_n(1,\cdots,n) \nonumber\\
&+ \left( (2+u_n\partial_{u_n}) \sum_{i=1}^{n-1} \frac{1}{z_i-z_n} \partial_{u_i} + \bar{\partial}_n \sum_{i=1}^{n-1} \frac{\bar{z}_i-\bar{z}_n}{z_i-z_n} \partial_{u_i}\right)C_n(1,\cdots,n) = 0 \, , \label{eq:Cgradiff2}
\end{align}
respectively. Similar to Eq.(\ref{eq:C3Mag}), we can write down a magnetic branch solution for Eqs.(\ref{eq:Cgradiff1}) and (\ref{eq:Cgradiff2}),
\be
C_3(--,--,++)_{\text{Mag}} = \frac{z_{12}^{5/2}}{ z_{13}^{3/2} z_{23}^{3/2}} \frac{\bar{z}_{23}^{1/2} \bar{z}_{13}^{1/2}}{\bar{z}_{12}^{7/2}} \, ,
\ee
which has the expected Carrollian weights and dimensions.

Similar to Eq.(\ref{eq:C3flat}), one can compute the Carrollian three-point graviton amplitudes in flat space,
\be
C_3(--,--,++)_{\text{flat}} = \frac{z_{12}^2}{z_{13}z_{23}} \, \frac{1}{u_3-\frac{z_{32}}{z_{12}} u_1 -\frac{z_{31}}{z_{21}}u_2} \, \delta(\bar{z}_{13})\delta(\bar{z}_{23}) \, , \label{eq:C3gra}
\ee
where particles 1 and 2 are incoming, particle 3 is outgoing. We have neglected an overall constant. One can check that Eq.(\ref{eq:C3gra}) satisfies both Eqs.(\ref{eq:Cgradiff1}) and (\ref{eq:Cgradiff2}) by using identities such as $\bar{z}_{23} \partial_{\bar{z}_2} \delta(\bar{z}_{23}) = -\delta(\bar{z}_{23})$ and $\bar{z}_{23}\,\delta(\bar{z}_{23})=0$.

A comment is in order about the $\overline{\text{MHV}}$ case. Similar to Eq.(\ref{eq:CarrollBGMHVbar}), from Eqs.(\ref{eq:Cgradiff1}) and (\ref{eq:Cgradiff2}), we can obtain differential equations satisfied by $\overline{\text{MHV}}$ Carrollian graviton amplitudes by switching $z$ and $\bar{z}$, and flipping $J$ to $-J$, 
\begin{align}
&-\sum_{i=1}^{n-1} \frac{(1+u_i\partial_{u_i}+J_i) ({z}_i-{z}_n) +({z}_i-{z}_n)^2{\partial}_i}{\bar{z}_i-\bar{z}_n} \partial_{u_n} C_{n,\overline{\text{MHV}}}(1,\cdots,n)  \nonumber\\
&+(2+u_n\partial_{u_n} +J_n) \sum_{i=1}^{n-1} \frac{{z}_i-{z}_n}{\bar{z}_i-\bar{z}_n} \partial_{u_i} C_{n,\overline{\text{MHV}}}(1,\cdots,n) =0 \, , \label{eq:Cgradiff1barmhv}
\end{align}
\begin{align}
&\left( \bar{\partial}_n \partial_{u_n} - \sum_{i=1}^{n-1} \frac{(1+u_i\partial_{u_i} +J_i) +2({z}_i-{z}_n) {\partial}_i}{\bar{z}_i-\bar{z}_n} \, \partial_{u_n}\right)C_{n,\overline{\text{MHV}}}(1,\cdots,n) \nonumber\\
&+ \left( (2+u_n\partial_{u_n}) \sum_{i=1}^{n-1} \frac{1}{\bar{z}_i-\bar{z}_n} \partial_{u_i} + {\partial}_n \sum_{i=1}^{n-1} \frac{{z}_i-{z}_n}{\bar{z}_i-\bar{z}_n} \partial_{u_i}\right)C_{n,\overline{\text{MHV}}}(1,\cdots,n) = 0 \, . \label{eq:Cgradiff2barmhv}
\end{align}

\subsection{From leading, subleading, and subsubleading soft graviton symmetries}
In \cite{Banerjee:2021cly}, the authors showed null state conditions constructed from leading, subleading, and subsubleading soft graviton symmetries,
\begin{equation}
    S^0_{-1}G_{\Delta}^{+2}(z,\bar{z})= \frac{1}{2} (\Delta-2)(\Delta-3) P_{-2,0} \, G^{+2}_{\Delta-2}(z,\bar{z}) \, , \label{eq:BGsubsubdiff1}
\end{equation}
\begin{equation}
    S_{-1}^1 \, G^{+2}_{\Delta}(z,\bar{z}) \, - \frac{1}{2} (\Delta-1)(\Delta-3) \, P_{-2,-1} \, G^{+2}_{\Delta-2}(z,\bar{z}) \, +\, 2(\Delta-2) J_{-1}^0 \, P_{-1,-1}\, \, G^{+2}_{\Delta-2}(z,\bar{z}) \, = 0 \, .
\end{equation}
In terms of celestial MHV graviton amplitudes, they lead to 
\begin{align}
&-\sum_{i=1}^{n-1} \frac{\bar{z}_i-\bar{z}_n}{z_i-z_n} \left( 2\bar{h}_i(2\bar{h}_i-1) +2(\bar{z}_i-\bar{z}_n) 2\bar{h}_i \bar{\partial}_i + (\bar{z}_i-\bar{z}_n)^2 \bar{\partial}_i^2\right) \epsilon_i e^{-\frac{\partial}{\partial\Delta_i}} \mathcal{M}_n \nonumber\\
&+ (\Delta_n-2)(\Delta_n-3) \sum_{i=1}^{n-1} \frac{\bar{z}_i-\bar{z}_n}{z_i-z_n} \epsilon_i e^{\frac{\partial}{\partial\Delta_i}} e^{-2 \frac{\partial}{\partial\Delta_n}} \mathcal{M}_n = 0 \, ,
\end{align}
\begin{align}
&\sum_{i=1}^{n-1} \frac{1}{z_i-z_n}\left( 2\bar{h}_i(2\bar{h}_i-1) +8 (\bar{z}_i-\bar{z}_n)\bar{h}_i \bar{\partial}_i+3(\bar{z}_i-\bar{z}_n)^2 \bar{\partial}_i\right)\epsilon_i e^{-\frac{\partial}{\partial\Delta_i}} \mathcal{M}_n \nonumber\\
&(\Delta_n-2)(\Delta_n-3) \sum_{i=1}^{n-1} \frac{1}{z_i-z_n} \epsilon_i \,e^{\frac{\partial}{\partial\Delta_i}} \, e^{-2\frac{\partial}{\partial\Delta_n}}\mathcal{M}_n -4(\Delta_n-2)\sum_{i=1}^{n-1} \frac{\bar{h}_i+(\bar{z}_i-\bar{z}_n)\bar{\partial_i}}{z_i-z} \, e^{-\frac{\partial}{\partial\Delta_n}} \mathcal{M}_n = 0 \, .
\end{align}
Translating them into differential equations of the corresponding Carrollian amplitudes, we find the following differential equations of $u-$ descendants for Carrollian amplitudes,
\begin{align}
&\sum_{i=1}^{n-1} \frac{\bar{z}_i-\bar{z}_n}{z_i-z_n} \left( (2-3J_i)+ (4-2J_i)u_i\partial_{u_i} +u_i^2 \partial^2_{u_i} + J_i^2\right)\prod_{k\neq i}^{n-1} \partial_{u_k} \partial_{u_n}^2 C_n \nonumber\\
+&\sum_{i=1}^{n-1} \frac{\bar{z}_i-\bar{z}_n}{z_i-z_n} \left( 2(\bar{z}_i-\bar{z}_n)\bar{\partial}_i(2+u_i\partial_{u_i}-J_i) +(\bar{z}_i-\bar{z}_n)^2\bar{\partial}_i^2\right) \prod_{k\neq i}^{n-1} \partial_{u_k} \partial_{u_n}^2 C_n  \nonumber\\
+&( 2 u_n \partial_{u_n} +u_n^2 \partial^2_{u_n}) \sum_{i=1}^{n-1} \frac{\bar{z}_i-\bar{z}_n}{z_i-z_n} \partial_{u_i} \prod_{k=1}^{n-1} \partial_{u_k} C_n= 0 \, ,
\end{align}
\begin{align}
&\sum_{i=1}^{n-1} \frac{1}{z_i-z_n} \left( (2-3J_i)+ (4-2J_i)u_i\partial_{u_i} +u_i^2 \partial^2_{u_i} + J_i^2\right)\prod_{k\neq i}^{n-1} \partial_{u_k} \partial_{u_n}^2 C_n \nonumber\\
+&\sum_{i=1}^{n-1} \frac{1}{z_i-z_n} \left( 4(\bar{z}_i-\bar{z}_n)\bar{\partial}_i(2+u_i\partial_{u_i}-J_i) +3(\bar{z}_i-\bar{z}_n)^2\bar{\partial}_i^2\right) \prod_{k\neq i}^{n-1} \partial_{u_k} \partial_{u_n}^2 C_n  \nonumber\\
+&( 2 u_n \partial_{u_n} +u_n^2 \partial^2_{u_n}) \sum_{i=1}^{n-1} \frac{1}{z_i-z_n} \partial_{u_i} \prod_{k=1}^{n-1} \partial_{u_k} C_n\nonumber\\
&- (2+2u_n\partial_{u_n}) \sum_{i=1}^{n-1} \frac{(2+u_i\partial_{u_i} -J_i)+(\bar{z}_i-\bar{z}_n)\bar{\partial}_i}{z_i-z_n} \prod_{k=1}^{n-1} \partial_{u_k} \partial_{u_n} C_n= 0 \, .
\end{align}

\section{Covariance of Carrollian differential equations} \label{sec5}

In this section, we show that the differential equations for MHV gluon amplitudes \eqref{eq:CarrollBG} and MHV graviton amplitudes \eqref{eq:Cgradiff1}-\eqref{eq:Cgradiff2} are genuine Carrollian equations. More precisely, we verify the compatibility of these equations with global conformal Carrollian transformations at null infinity, provided the correlators satisfy the Carrollian Ward identities.
Of course, this can be understood as a manifestation of Poincar\'e invariance in the bulk since the global conformal algebra in three dimensions is isomorphic to the Poincar\'e algebra in four dimensions (see e.g. Appendix B of \cite{Donnay:2022wvx}). Nevertheless, these equations are of interest for Carrollian physics and it is instructive to check the invariance from an intrinsic boundary perspective. 

The finite global conformal Carrollian transformations acting on the coordinates $x=(u,z,\bar{z})$ at $\mathscr{I}$ are given by
\begin{equation}
u' = \frac{1}{|cz+d|^2} [u +  T(z, \bar{z}) ], \qquad z' = \frac{a z + b}{cz + d}, \qquad \bar{z}' = \frac{\bar a \bar z + \bar b}{\bar c \bar z + \bar d}
\label{BMS coord transformation}
\end{equation} Here $a,b,c,d \in \mathbb{C}$ satisfy $ad-bc = 1$ parametrize the six $SL(2, \mathbb{C})$ transformations, and $T(z,\bar{z}) = \sum_{0 \le m,n \le 2} t_{m,n} z^m \bar{z}^n$ with $t_{m,n}^*= t_{n,m} \in \mathbb{C}$ parametrize the four (bulk) translations. In particular, the two Carrollian boosts in three dimensions are generated by $T = z, \bar{z}$. The Carrollian amplitudes $C_n$ in \eqref{eq:nCarrollian} have been shown to satisfy Carrollian Ward identities \cite{Donnay:2022wvx,Mason:2023mti}. The finite version of these is given by
\begin{equation}
C_n'(x_1' \ldots x_n') = \Big( \frac{\partial z}{\partial z'} \Big)_{z=z_1}^{k_1} \Big( \frac{\partial \bar z}{\partial \bar z'} \Big)_{\bar z= \bar z_1}^{\bar k_1} \ldots  \Big( \frac{\partial z}{\partial z'} \Big)_{z=z_n}^{k_n} \Big( \frac{\partial \bar z}{\partial \bar z'} \Big)_{\bar z= \bar z_n}^{\bar k_n} C_n (x_1 \ldots x_n)
\label{Carrollian WI}
\end{equation} where the Carrollian weights $(k,\bar k_i)$ are fixed in terms of the particle helicities $J_i$ as \eqref{carrollian weights}. This justifies the identification of position space amplitudes at $\mathscr{I}$ with Carrollian CFT correlators in \eqref{holographic identification}.  Crucially, Carrollian boost Ward identities imply the division of the correlators into electric and magnetic branches, as discussed around \eqref{elec vs magn} (see also \cite{Chen:2021xkw,Donnay:2022wvx}). 

Invariance of the differential equations \eqref{eq:CarrollBG} and \eqref{eq:Cgradiff1}-\eqref{eq:Cgradiff2} under $SL(2,\mathbb{C})$ is guaranteed from the starting point given by the BG equations in celestial CFT \eqref{eq:nptBG} and \eqref{eq:Mgradiff1}-\eqref{eq:Mgradiff2}. In particular, dilatation invariance can be checked directly from a simple weights counting. From \eqref{BMS coord transformation}, one can easily deduce
\be
z_{ij}\bar{z}_{ij} \rightarrow \frac{z_{ij}\bar{z}_{ij}}{(c z_i+d)(\bar{c}\bar{z}_i+\bar{d})(c z_j+d)(\bar{c}\bar{z}_j+\bar{d})} \, .
\ee
so that the following combinations are $SL(2,\mathbb{C})$ invariant:
\be
d_{ij} = \frac{z_{ij}\bar{z}_{ij}}{u_iu_j} \, , \qquad   e_{ijk} = \frac{u_i z_{jk}\bar{z}_{jk} - u_j z_{ki}\bar{z}_{ki}}{u_i z_{jk}\bar{z}_{jk} + u_j z_{ki}\bar{z}_{ki}}      \, .
\ee
These objects $d_{ij}$ and $e_{ijk}$ appear naturally in the non-distributional solutions of Carrollian BG equations discussed in the previous sections. 

Here we focus on (bulk) translations, which include the (boundary) Carrollian boosts. Under translation, the derivative operators transform according to the chain rule:
\begin{equation}
\partial_{u'} = \partial_u , \qquad \partial_{z'} = \partial_z - \partial_z T \partial_u, \qquad  \partial_{\bar z'} = \partial_{\bar z} - \partial_{\bar z} T \partial_u
\end{equation} Using this together with \eqref{Carrollian WI}, we show explicitly that \eqref{eq:CarrollBG} and \eqref{eq:Cgradiff1}-\eqref{eq:Cgradiff2} are translation invariant.  

\paragraph{Differential equation for gluons}   Denoting by $\mathcal{G}_{\text{gluons}}(x_1 \ldots x_n)$ the left-hand side of  \eqref{eq:CarrollBG}, transformation under translations yields
\begin{equation}
\begin{split}
&\mathcal{G}_{\text{gluons}}(x'_1 \ldots x'_n) \\
&=  \left( \partial_i' -\frac{1+u'_i\partial_{u_i'}}{z'_{i-1,i}} -\frac{1}{z'_{i+1,i}}\right)\partial_{u'_{i-1}}C_n(x_1' \cdots x_n') \\
&+ \left(  \frac{1+u'_{i-1}\partial_{u'_{i-1}} -J_{i-1}+\bar{z}'_{i-1,i} \, \bar{\partial}'_{i-1}}{z'_{i-1,i}} \right) \partial_{u'_{i}} C_n(x_1' \cdots x_n')  \\
&= \left( \partial_i -\frac{1+u_i\partial_{u_i}}{z_{i-1,i}} -\frac{1}{z_{i+1,i}}\right)\partial_{u_{i-1}}C_n(x_1 \cdots x_n) \\
&+ \left(  \frac{1+u_{i-1}\partial_{u_{i-1}} -J_{i-1}+\bar{z}_{i-1,i} \, \bar{\partial}_{i-1}}{z_{i-1,i}} \right) \partial_{u_{i}} C_n(x_1 \cdots x_n) \\
&+\Big[-\partial_i T(z_i, \bar{z}_i) - \frac{1}{z_{i-1,i}} T(z_i, \bar{z}_i)+ \frac{1}{z_{i-1,i}} T(z_{i-1}, \bar{z}_{i-1}) \\
&\qquad\qquad\qquad\qquad\qquad - \frac{\bar{z}_{i-1,i}}{z_{i-1,i}} \bar{\partial}_{i-1} T(z_{i-1},\bar{z}_{i-1}) \Big] \partial_{u_i-1} \partial_{u_i} C_n (x_1 \cdots x_n)
\end{split}
\end{equation} The two last lines can be shown to cancel each other for the four translation generators $T(z,\bar{z}) = 1, z, \bar{z}, z \bar{z}$. 

\paragraph{Differential equations for gravitons} Similarly, we denote by $\mathcal{G}_{\text{gravitons}}^{(1)}(x_1 \ldots x_n)$ and $\mathcal{G}_{\text{gravitons}}^{(2)}(x_1 \ldots x_n)$ the left-hand sides of \eqref{eq:Cgradiff1} and \eqref{eq:Cgradiff2}, respectively. Transformation under translations implies
\begin{equation}
\begin{split}
&\mathcal{G}_{\text{gravitons}}^{(1)}(x'_1 \ldots x'_n) = \mathcal{G}_{\text{gravitons}}^{(1)}(x_1 \ldots x_n)  \\
&\qquad\qquad+ \sum_{i=1}^{n-1} \frac{\bar{z}_{i,n}}{z_{i,n}}\Big[ -  T(z_i,\bar{z}_i) + \bar{z}_{i,n} \bar{\partial}_i T(z_i,\bar{z}_i) + T(z_n,\bar{z}_n )  \Big] \partial_{u_i}\partial_{u_n} C_n(x_1 \ldots x_n)
\end{split}
\end{equation} One can check that the last line vanishes immediately for holomorphic translations $T = 1, z$. For $T = \bar{z}, z\bar{z}$, the last line can be shown to vanish by noticing that
\begin{equation}
\begin{split}
\sum_{i = 1}^{n-1} \bar{z}_{i,n} \partial_{u_i} \partial_{u_n} C_n &= \sum_{i = 1}^{n-1} \bar{z}_{i} \partial_{u_i} \partial_{u_n} C_n  - \bar{z}_{n} \sum_{i = 1}^{n-1}  \partial_{u_i} \partial_{u_n} C_n  \\
&= \sum_{i = 1}^{n-1} \bar{z}_{i} \partial_{u_i} \partial_{u_n} C_n  + \bar{z}_{n} \partial_{u_n} \partial_{u_n} C_n \\
&= \sum_{i = 1}^{n} \bar{z}_{i} \partial_{u_i} \partial_{u_n} C_n \\
&=0
\end{split} 
\end{equation} where we used the Ward identities for translations in the second and last equalities. Similarly, the transformation of \eqref{eq:Cgradiff2} under translations leads to 
\begin{equation}
\begin{split}
&\mathcal{G}_{\text{gravitons}}^{(2)}(x'_1 \ldots x'_n) = \mathcal{G}_{\text{gravitons}}^{(2)}(x_1 \ldots x_n) - \partial_n T (z_n, \bar{z}_n) \partial_{u_n}^2 C_n \\
&+ \sum_{i=1}^{n-1} \frac{1}{z_{i,n}}\Big[ T(z_n,\bar{z}_n) - T(z_i, \bar{z}_i)+ 2 \bar{z}_{i,n} \bar{\partial}_i T (z_i, \bar{z}_i) - \bar{\partial}_n T(z_n,\bar{z}_n) \bar{z}_{i,n} \Big] \partial_{u_i}\partial_{u_n} C_n(x_1 \ldots x_n)
\end{split}
\end{equation} The extra terms cancel immediately for $T = 1, \bar{z}$, and can be shown to cancel for $T = z, z\bar{z}$ after using the Ward identities for translations. 

Hence, we conclude that the Carrollian BG equations are translation invariant. Of course, as discussed in the previous sections, these equations also admit solutions which explicitly break translation invariance. Acting with translations on these solutions provides a systematic way to produce new solutions.

\section{Concluding remarks} \label{sec6}


In \cite{Banerjee:2020zlg,Banerjee:2020vnt}, it was shown that the celestial differential equations impose constraints on the celestial OPEs.
It would be interesting to check if our Carrollian differential equations impose non-trivial constraints on the Carrollian OPEs and compare them with the OPEs shown in \cite{Mason:2023mti}.
It would be also interesting to see if there are alternative ways of deriving our Carrollian differential equations, perhaps starting from finding some Carrollian null state conditions.
We expect Eqs.(\ref{eq:granull1}) and (\ref{eq:granull2}) would be translated to null state conditions in 3D Carrollian CFT. A related approach would be to take the Carrollian limit of the differential equations in 3D CFT, which, in the current holographic framework, could be understood from a flat limit in the bulk, as discussed in \cite{Alday:2024yyj}. However, the relation between null state conditions in 3D CFTs and differential equations of correlation functions has been largely unexplored, due to the fact that the usual three-dimensional conformal group is finite-dimensional, and lack of the holomorphic and anti-holomorphic factorization in $d>2$.\footnote{See however, recent work in \cite{Huang:2023ikg,Huang:2024wbq} searching for BPZ-type equations in CFTs in $d>2$.}
For the celestial case, using the null state condition shown in \cite{Banerjee:2020zlg,Banerjee:2021cly}, we show a sample calculation of constructing a celestial null state from subleading and subsubleading soft graviton symmetries in Appendix \ref{secAppA}, which would hopefully shed some lights towards this direction.

We have shown that the differential equations for Carrollian amplitudes are Carrollian covariant. It would be interesting to find a manifestly covariant form of these equations using some Carrollian geometry at $\mathscr{I}$. It would also be fascinating to see whether these equations can be derived from a variational principle which would provide an example of Carrollian CFT action relevant for holography. 

It would be interesting to find a physical interpretation of the magnetic solutions of the Carrollian differential equations shown in Sections \ref{sec3} and \ref{sec4}. As discussed around Eq.\eqref{inverse fourier}, one might expect them to be related to soft modes or Goldstones modes.

Recently, celestial holography on non-trivial backgrounds has attracted considerable attention \cite{Fan:2022vbz,Casali:2022fro,Fan:2022kpp,deGioia:2022fcn,Gonzo:2022tjm,Pasterski:2020pdk,Banerjee:2023rni,Ball:2023ukj,Crawley:2023brz,Costello:2022wso,Melton:2022fsf,Bittleston:2023bzp,Costello:2022jpg,Costello:2023hmi,Adamo:2023zeh,Adamo:2024mqn,Melton:2023dee,Taylor:2023ajd,Bittleston:2024rqe,Adamo:2024xpc}. Most of the work focuses on using backgrounds to regularize the singularities of celestial amplitudes and the fate of celestial soft algebras upon turning on backgrounds.
It would be interesting to explore whether there is a similar story for the Carrollian case. Our results in Section \ref{sec2} suggest  encouraging steps towards this direction.   
Instead of massless dilaton backgrounds that we considered in Section \ref{sec2}, one can definitely consider massive dilaton backgrounds. For the celestial case, massive dilaton backgrounds have been used in \cite{Casali:2022fro,Banerjee:2023rni,Ball:2023ukj,Taylor:2023bzj}.

Solutions of the Knizhnik--Zamolodchikov (KZ) equation describe correlation functions of primary (or descendant) fields  in two--dimensional CFTs.
Furthermore, the Drinfeld associator  captures  the monodromy of a universal version of the KZ equation, which in turn naturally appears in the world--sheet monodromy representation  of  open superstring amplitudes  \cite{Broedel:2013aza}. Interestingly, the coefficients of the Drinfeld associator govern the Carrollian four--point open superstring amplitude \cite{Stieberger:2024shv}. It would be interesting  to understand whether our differential equation \req{eq:CarrollBG} corresponds to some variant of  KZ type equations of three--dimensional  Carrollian CFT.

\section*{Acknowledgements}
We would like to thank Shamik Banerjee and Partha Paul for useful correspondence. BZ would like to thank Tim Adamo, Yangrui Hu, and Gerben Oling for discussions on related topics. RR
is supported by the Titchmarsh Research Fellowship at
the Mathematical Institute and by the Walker Early Career Fellowship in Mathematical Physics at Balliol College. RR and StSt are grateful to the Erwin Schr\"odinger Institute in Vienna for  invitation to the workshop {\it Carrollian Physics and Holography}, where this project has been initiated. TRT is supported by the National Science Foundation
under Grants Number PHY-1913328 and PHY-2209903, by the 
NAWA Grant 
``Celestial Holography of Fundamental Interactions'' and 
by the Simons Collaboration on Celestial Holography.
Any opinions, findings, and conclusions or
recommendations expressed in this material are those of the authors and do not necessarily
reflect the views of the National Science Foundation. BZ is supported by Royal Society.

\appendix 
\section{A differential equation of celestial MHV graviton amplitudes} \label{secAppA}
In \cite{Banerjee:2020zlg}, the authors showed a null state condition constructed from leading and subleading soft graviton symmetries,
\begin{equation}
   ( J^1_{-1}P_{-1,-1}-(\Delta-1) P_{-2,0})\, G_{\Delta}^{+2}(z,\bar{z}) = 0 \, ,
\label{null state condition}
\end{equation}
where $J^1_{-1}$ is associated to subleading soft graviton.

In  \cite{Banerjee:2021cly}, the authors showed another null state condition constructed from leading, and subsubleading soft graviton symmetries,
\begin{equation}
    S^0_{-1}G_{\Delta}^{+2}(z,\bar{z})= \frac{1}{2} (\Delta-2)(\Delta-3) P_{-2,0} \, G^{+2}_{\Delta-2}(z,\bar{z}) \, , \label{eq:BGdiff2}
\end{equation}
where $S^0_{-1}$ is associated to sub-subleading soft graviton.

One can combine these two equations and eliminate $P_{-2,0}$ in the equation. We find an equation which is a linear combination of subleading and sub-subleading current algebra descendants only 
\begin{equation}
     \frac{1}{2}(\Delta-2)J^1_{-1}G^{+2}_{\Delta-1}(z,\bar{z}) = S^0_{-1}G^{+2}_{\Delta}(z,\bar{z}) \, . \label{eq:nullsubsub}
\end{equation}
Here we show an alternative way of deriving Eq.(\ref{eq:nullsubsub}) from the algebras among $J$ and $S$ without using $P$.

Eq.(A.1) in \cite{Banerjee:2021cly} tells us that $S^0_{-1}$ acting on a primary $\phi_{h,\bar{h}}$ with conformal weights $(h,\bar{h})$ gives us a descendant with weights $(h+\frac{1}{2}, \bar{h}-\frac{3}{2})$. Similarly $J^1_{-1}\,  \phi_{h-\frac{1}{2},\bar{h}-\frac{1}{2}}$ is also a descendant with weights $(h+\frac{1}{2}, \, \bar{h}-\frac{3}{2})$. So the null state condition we are looking for takes the following form
\begin{equation}
    A(\Delta) J^{1}_{-1} \, G^{+2}_{\Delta-1} = S^{0}_{-1} G^{+2}_{\Delta} \, . \label{eq:nsubsub1}
\end{equation}
We will use the algebra of $J$ and $S$ to find $A(\Delta)$. We use Eq.(A.10) in \cite{Banerjee:2021cly}:
\begin{equation}
   [ S^1_{0}, J^{1}_{-1}] = 3 S^0_{-1} \, ,
\end{equation}
 and Eq.(A.14) in \cite{Banerjee:2021cly}:
\begin{equation}
    [S^1_0, S^0_{-1}] \phi_{h,\bar{h}}  = 4 \bar{h} \,  S^0_{-1} \, \phi_{h-\frac{1}{2},\bar{h}-\frac{1}{2}} \, ,
\end{equation}
We compute the commutator between $S^1_0$ and (\ref{eq:nsubsub1}),
\begin{equation}
    \begin{split}
        &A(\Delta)[S^1_0, J^1_{-1} G^{+2}_{\Delta-1}]-[S^1_0, S^0_{-1}G^{+2}_{\Delta}]= \\
        &=A(\Delta) \left(3 S^0_{-1}G^{+2}_{\Delta-1}+J^{1}_{-1}[S^1_0,G^{+2}_{\Delta-1}]\right)-4 \frac{\Delta-2}{2} S^0_{-1} G^{+2}_{\Delta-1} - S^0_{-1}[S^1_{0},G^{+2}_{\Delta}] = 0 \, .
    \end{split}
\end{equation}
Using Eq.(5.3) in \cite{Banerjee:2021cly}:
\begin{equation}
    [S^1_0, G^{+2}_{\Delta}] = -\frac{1}{2}(\Delta-2)(\Delta-3) G^{+2}_{\Delta-1} \, 
\end{equation}
and also (\ref{eq:nsubsub1}) to eliminate $J^1_{-1}$, we obtain an equation for $A(\Delta)$
\begin{equation}
    3A(\Delta)-\frac{1}{2}(\Delta-3)(\Delta-4) \frac{A(\Delta)}{A(\Delta-1)} = 2(\Delta-2)-\frac{1}{2}(\Delta-2)(\Delta-3) \, .
\end{equation}
One can check $A(\Delta) = \frac{1}{2}(\Delta-2)$ is the solution, which agrees with Eq.(\ref{eq:nullsubsub}).
Inserting (\ref{eq:nullsubsub}) into celestial MHV graviton amplitudes, we find the following differential equation
\begin{equation}
    \frac{1}{2}(\Delta-2) \mathcal{J}^{1}_{-1}  \, e^{-\partial_{\Delta}} \, \langle G^{+2}_{\Delta}(z,\bar{z}) \prod_i G^{\sigma_i}_{\Delta_i}(z_i,\bar{z}_i) \rangle_{MHV}  = \mathcal{S}^0_{-1} \langle G^{+2}_{\Delta}(z,\bar{z}) \prod_i G^{\sigma_i}_{\Delta_i}(z_i,\bar{z}_i) \rangle_{MHV} \, , \label{eq:diff28}
\end{equation}
where
\begin{equation}
    \mathcal{J}^1_{-1} = -\sum_i \frac{2\bar{h}_i (\bar{z}_i-\bar{z})+(\bar{z}_i-\bar{z})^2 \bar{\partial}_i}{z_i-z} \, , \label{eq:J1-1}
\end{equation}
\begin{equation}
    \mathcal{S}^0_{-1} = -\frac{1}{2} \sum_{i} \frac{\bar{z}_i-\bar{z}}{z_i-z}\Big[ 2\bar{h}_i(2\bar{h}_i-1)+2(\bar{z_i}-\bar{z}) 2\bar{h}_i \bar{\partial}_i + (\bar{z}_i-\bar{z})^2 \bar{\partial}_i^2 \Big]\, \epsilon_i \, e^{-\partial_{\Delta_i}}\, . \label{eq:S0-1}
\end{equation}

Given the fact that now we have a differential equation constructed from subleading and subsubleading soft graviton symmetries, which are independent of the leading soft graviton symmetries related to translation symmmetries in the bulk, one can show that Eq.(\ref{eq:diff28})  admit the following three-point non-distributional solution,
\begin{align}
&\mathcal{M}_3(\Delta_1=-2+i\lambda_1,J_1=-2,\, \Delta_2=-2+i\lambda_2,J_2=-2, \, \Delta_3=2+i\lambda_3, J_3=2) \nonumber\\
=&\delta(\lambda_1+\lambda_2+\lambda_3)B(1-i\lambda_1,1-i\lambda_2)\frac{z_{12}^6}{z_{13}^2z_{23}^2} (z_{12}\bar{z}_{12})^{i\lambda_3}(z_{23}\bar{z}_{23})^{i\lambda_1}(z_{13}\bar{z}_{13})^{i\lambda_2} \, ,
\end{align}
where the incoming/outgoing configuration has been chosen to be :$\epsilon_1=\epsilon_2=-1,\epsilon_3=1$.
It would be interesting to see if one can compute it from other prescriptions, perhaps along the line with \cite{Ball:2023ukj}.


\begin{thebibliography}{99}

\bibitem{Belavin:1984vu}
A.~A.~Belavin, A.~M.~Polyakov and A.~B.~Zamolodchikov,
``Infinite Conformal Symmetry in Two-Dimensional Quantum Field Theory,''
Nucl. Phys. B \textbf{241}, 333-380 (1984)
doi:10.1016/0550-3213(84)90052-X

\bibitem{Knizhnik:1984nr}
V.~G.~Knizhnik and A.~B.~Zamolodchikov,
``Current Algebra and Wess-Zumino Model in Two-Dimensions,''
Nucl. Phys. B \textbf{247}, 83-103 (1984)
doi:10.1016/0550-3213(84)90374-2

\bibitem{Dolan:2011dv}
F.~A.~Dolan and H.~Osborn,
``Conformal Partial Waves: Further Mathematical Results,''
[arXiv:1108.6194 [hep-th]].

\bibitem{DiFrancesco:1997nk}
P.~Di Francesco, P.~Mathieu and D.~Senechal,
``Conformal Field Theory,''
Springer-Verlag, 1997,
ISBN 978-0-387-94785-3, 978-1-4612-7475-9
doi:10.1007/978-1-4612-2256-9

\bibitem{Pasterski:2016qvg}
S.~Pasterski, S.~H.~Shao and A.~Strominger,
``Flat Space Amplitudes and Conformal Symmetry of the Celestial Sphere,''
Phys. Rev. D \textbf{96}, no.6, 065026 (2017)
doi:10.1103/PhysRevD.96.065026
[arXiv:1701.00049 [hep-th]].

\bibitem{Pasterski:2017kqt}
S.~Pasterski and S.~H.~Shao,
``Conformal basis for flat space amplitudes,''
Phys. Rev. D \textbf{96}, no.6, 065022 (2017)
doi:10.1103/PhysRevD.96.065022
[arXiv:1705.01027 [hep-th]].

\bibitem{Pasterski:2017ylz}
S.~Pasterski, S.~H.~Shao and A.~Strominger,
``Gluon Amplitudes as 2d Conformal Correlators,''
Phys. Rev. D \textbf{96}, no.8, 085006 (2017)
doi:10.1103/PhysRevD.96.085006
[arXiv:1706.03917 [hep-th]].


\bibitem{Stieberger:2018onx}
S.~Stieberger and T.~R.~Taylor,
``Symmetries of Celestial Amplitudes,''
Phys. Lett. B \textbf{793}, 141-143 (2019)
doi:10.1016/j.physletb.2019.03.063
[arXiv:1812.01080 [hep-th]].

\bibitem{Arkani-Hamed:2020gyp}
N.~Arkani-Hamed, M.~Pate, A.~M.~Raclariu and A.~Strominger,
``Celestial amplitudes from UV to IR,''
JHEP \textbf{08}, 062 (2021)
doi:10.1007/JHEP08(2021)062
[arXiv:2012.04208 [hep-th]].

\bibitem{Donnay:2018neh}
L.~Donnay, A.~Puhm and A.~Strominger,
``Conformally Soft Photons and Gravitons,''
JHEP \textbf{01}, 184 (2019)
doi:10.1007/JHEP01(2019)184
[arXiv:1810.05219 [hep-th]].

\bibitem{Fan:2019emx}
W.~Fan, A.~Fotopoulos and T.~R.~Taylor,
``Soft Limits of Yang-Mills Amplitudes and Conformal Correlators,''
JHEP \textbf{05}, 121 (2019)
doi:10.1007/JHEP05(2019)121
[arXiv:1903.01676 [hep-th]].

\bibitem{Pate:2019mfs}
M.~Pate, A.~M.~Raclariu and A.~Strominger,
``Conformally Soft Theorem in Gauge Theory,''
Phys. Rev. D \textbf{100}, no.8, 085017 (2019)
doi:10.1103/PhysRevD.100.085017
[arXiv:1904.10831 [hep-th]].

\bibitem{Adamo:2019ipt}
T.~Adamo, L.~Mason and A.~Sharma,
``Celestial amplitudes and conformal soft theorems,''
Class. Quant. Grav. \textbf{36}, no.20, 205018 (2019)
doi:10.1088/1361-6382/ab42ce
[arXiv:1905.09224 [hep-th]].

\bibitem{Puhm:2019zbl}
A.~Puhm,
``Conformally Soft Theorem in Gravity,''
JHEP \textbf{09}, 130 (2020)
doi:10.1007/JHEP09(2020)130
[arXiv:1905.09799 [hep-th]].

\bibitem{Guevara:2019ypd}
A.~Guevara,
``Notes on Conformal Soft Theorems and Recursion Relations in Gravity,''
[arXiv:1906.07810 [hep-th]].


\bibitem{Pate:2019lpp}
M.~Pate, A.~M.~Raclariu, A.~Strominger and E.~Y.~Yuan,
``Celestial operator products of gluons and gravitons,''
Rev. Math. Phys. \textbf{33}, no.09, 2140003 (2021)
doi:10.1142/S0129055X21400031
[arXiv:1910.07424 [hep-th]].

\bibitem{Fotopoulos:2019vac}
A.~Fotopoulos, S.~Stieberger, T.~R.~Taylor and B.~Zhu,
``Extended BMS Algebra of Celestial CFT,''
JHEP \textbf{03}, 130 (2020)
doi:10.1007/JHEP03(2020)130
[arXiv:1912.10973 [hep-th]].

\bibitem{Banerjee:2020kaa}
S.~Banerjee, S.~Ghosh and R.~Gonzo,
``BMS symmetry of celestial OPE,''
JHEP \textbf{04}, 130 (2020)
doi:10.1007/JHEP04(2020)130
[arXiv:2002.00975 [hep-th]].

\bibitem{Donnay:2020guq}
L.~Donnay, S.~Pasterski and A.~Puhm,
``Asymptotic Symmetries and Celestial CFT,''
JHEP \textbf{09}, 176 (2020)
doi:10.1007/JHEP09(2020)176
[arXiv:2005.08990 [hep-th]].

\bibitem{Guevara:2021abz}
A.~Guevara, E.~Himwich, M.~Pate and A.~Strominger,
``Holographic symmetry algebras for gauge theory and gravity,''
JHEP \textbf{11}, 152 (2021)
doi:10.1007/JHEP11(2021)152
[arXiv:2103.03961 [hep-th]].

\bibitem{Strominger:2021mtt}
A.~Strominger,
``$w_{1+\infty}$ Algebra and the Celestial Sphere: Infinite Towers of Soft Graviton, Photon, and Gluon Symmetries,''
Phys. Rev. Lett. \textbf{127}, no.22, 221601 (2021)
doi:10.1103/PhysRevLett.127.221601
[arXiv:2105.14346 [hep-th]].

\bibitem{Himwich:2021dau}
E.~Himwich, M.~Pate and K.~Singh,
``Celestial operator product expansions and w$_{1+\infty}$ symmetry for all spins,''
JHEP \textbf{01}, 080 (2022)
doi:10.1007/JHEP01(2022)080
[arXiv:2108.07763 [hep-th]].



\bibitem{Banerjee:2020zlg}
S.~Banerjee, S.~Ghosh and P.~Paul,
``MHV graviton scattering amplitudes and current algebra on the celestial sphere,''
JHEP \textbf{02}, 176 (2021)
doi:10.1007/JHEP02(2021)176
[arXiv:2008.04330 [hep-th]].

\bibitem{Banerjee:2020vnt}
S.~Banerjee and S.~Ghosh,
``MHV gluon scattering amplitudes from celestial current algebras,''
JHEP \textbf{10}, 111 (2021)
doi:10.1007/JHEP10(2021)111
[arXiv:2011.00017 [hep-th]].


\bibitem{Banerjee:2021cly}
S.~Banerjee, S.~Ghosh and S.~S.~Samal,
``Subsubleading soft graviton symmetry and MHV graviton scattering amplitudes,''
JHEP \textbf{08}, 067 (2021)
doi:10.1007/JHEP08(2021)067
[arXiv:2104.02546 [hep-th]].

\bibitem{Pasterski:2021fjn}
S.~Pasterski, A.~Puhm and E.~Trevisani,
``Celestial diamonds: conformal multiplets in celestial CFT,''
JHEP \textbf{11}, 072 (2021)
doi:10.1007/JHEP11(2021)072
[arXiv:2105.03516 [hep-th]].

\bibitem{Hu:2021lrx}
Y.~Hu, L.~Ren, A.~Y.~Srikant and A.~Volovich,
``Celestial dual superconformal symmetry, MHV amplitudes and differential equations,''
JHEP \textbf{12}, 171 (2021)
doi:10.1007/JHEP12(2021)171
[arXiv:2106.16111 [hep-th]].


\bibitem{Hu:2022bpa}
Y.~Hu and S.~Pasterski,
``Celestial recursion,''
JHEP \textbf{01}, 151 (2023)
doi:10.1007/JHEP01(2023)151
[arXiv:2208.11635 [hep-th]].

\bibitem{Adamo:2022wjo}
T.~Adamo, W.~Bu, E.~Casali and A.~Sharma,
``All-order celestial OPE in the MHV sector,''
JHEP \textbf{03}, 252 (2023)
doi:10.1007/JHEP03(2023)252
[arXiv:2211.17124 [hep-th]].

\bibitem{Banerjee:2023zip}
S.~Banerjee, H.~Kulkarni and P.~Paul,
``An infinite family of w$_{1+\infty}$ invariant theories on the celestial sphere,''
JHEP \textbf{05}, 063 (2023)
doi:10.1007/JHEP05(2023)063
[arXiv:2301.13225 [hep-th]].

\bibitem{Saha:2023hsl}
A.~Saha,
``Carrollian approach to 1 + 3D flat holography,''
JHEP \textbf{06}, 051 (2023)
doi:10.1007/JHEP06(2023)051
[arXiv:2304.02696 [hep-th]].

\bibitem{Banerjee:2023bni}
S.~Banerjee, R.~Mandal, S.~Misra, S.~Panda and P.~Paul,
``All $S$ invariant gluon OPEs on the celestial sphere,''
[arXiv:2311.16796 [hep-th]].


\bibitem{Donnay:2022aba}
L.~Donnay, A.~Fiorucci, Y.~Herfray and R.~Ruzziconi,
``Carrollian Perspective on Celestial Holography,''
Phys. Rev. Lett. \textbf{129}, no.7, 071602 (2022)
doi:10.1103/PhysRevLett.129.071602
[arXiv:2202.04702 [hep-th]].

\bibitem{Bagchi:2022emh}
A.~Bagchi, S.~Banerjee, R.~Basu and S.~Dutta,
``Scattering Amplitudes: Celestial and Carrollian,''
Phys. Rev. Lett. \textbf{128}, no.24, 241601 (2022)
doi:10.1103/PhysRevLett.128.241601
[arXiv:2202.08438 [hep-th]].

\bibitem{Donnay:2022wvx}
L.~Donnay, A.~Fiorucci, Y.~Herfray and R.~Ruzziconi,
``Bridging Carrollian and celestial holography,''
Phys. Rev. D \textbf{107}, no.12, 126027 (2023)
doi:10.1103/PhysRevD.107.126027
[arXiv:2212.12553 [hep-th]].


\bibitem{Barnich:2012rz}
G.~Barnich, A.~Gomberoff and H.~A.~Gonz\'alez,
``Three-dimensional Bondi-Metzner-Sachs invariant two-dimensional field theories as the flat limit of Liouville theory,''
Phys. Rev. D \textbf{87} (2013) no.12, 124032
doi:10.1103/PhysRevD.87.124032
[arXiv:1210.0731 [hep-th]].



\bibitem{Bagchi:2019xfx}
A.~Bagchi, A.~Mehra and P.~Nandi,
``Field Theories with Conformal Carrollian Symmetry,''
JHEP \textbf{05} (2019), 108
doi:10.1007/JHEP05(2019)108
[arXiv:1901.10147 [hep-th]].


\bibitem{Henneaux:2021yzg}
M.~Henneaux and P.~Salgado-Rebolledo,
``Carroll contractions of Lorentz-invariant theories,''
JHEP \textbf{11} (2021), 180
doi:10.1007/JHEP11(2021)180
[arXiv:2109.06708 [hep-th]].


\bibitem{deBoer:2021jej}
J.~de Boer, J.~Hartong, N.~A.~Obers, W.~Sybesma and S.~Vandoren,
``Carroll Symmetry, Dark Energy and Inflation,''
Front. in Phys. \textbf{10} (2022), 810405
doi:10.3389/fphy.2022.810405
[arXiv:2110.02319 [hep-th]].

\bibitem{Chen:2023pqf}
B.~Chen, R.~Liu, H.~Sun and Y.~f.~Zheng,
``Constructing Carrollian field theories from null reduction,''
JHEP \textbf{11} (2023), 170
doi:10.1007/JHEP11(2023)170
[arXiv:2301.06011 [hep-th]].

\bibitem{Alday:2024yyj}
L.~F.~Alday, M.~Nocchi, R.~Ruzziconi and A.~Yelleshpur Srikant,
``Carrollian Amplitudes from Holographic Correlators,''
[arXiv:2406.19343 [hep-th]].


\bibitem{Salzer:2023jqv}
J.~Salzer,
``An embedding space approach to Carrollian CFT correlators for flat space holography,''
JHEP \textbf{10}, 084 (2023)
doi:10.1007/JHEP10(2023)084
[arXiv:2304.08292 [hep-th]].

\bibitem{Saha:2023abr}
A.~Saha,
``w$_{1+\infty}$ and Carrollian holography,''
JHEP \textbf{05}, 145 (2024)
doi:10.1007/JHEP05(2024)145
[arXiv:2308.03673 [hep-th]].

\bibitem{Nguyen:2023miw}
K.~Nguyen,
``Carrollian conformal correlators and massless scattering amplitudes,''
JHEP \textbf{01}, 076 (2024)
doi:10.1007/JHEP01(2024)076
[arXiv:2311.09869 [hep-th]].

\bibitem{Bagchi:2023cen}
A.~Bagchi, P.~Dhivakar and S.~Dutta,
``Holography in Flat Spacetimes: the case for Carroll,''
[arXiv:2311.11246 [hep-th]].

\bibitem{Mason:2023mti}
L.~Mason, R.~Ruzziconi and A.~Yelleshpur Srikant,
``Carrollian Amplitudes and Celestial Symmetries,''
[arXiv:2312.10138 [hep-th]].

\bibitem{Liu:2024nfc}
W.~B.~Liu, J.~Long and X.~Q.~Ye,
``Feynman rules and loop structure of Carrollian amplitudes,''
JHEP \textbf{05}, 213 (2024)
doi:10.1007/JHEP05(2024)213
[arXiv:2402.04120 [hep-th]].

\bibitem{Have:2024dff}
E.~Have, K.~Nguyen, S.~Prohazka and J.~Salzer,
``Massive carrollian fields at timelike infinity,''
[arXiv:2402.05190 [hep-th]].

\bibitem{Stieberger:2024shv}
S.~Stieberger, T.~R.~Taylor and B.~Zhu,
``Carrollian Amplitudes from Strings,''
JHEP \textbf{04}, 127 (2024)
doi:10.1007/JHEP04(2024)127
[arXiv:2402.14062 [hep-th]].

\bibitem{Adamo:2024mqn}
T.~Adamo, W.~Bu, P.~Tourkine and B.~Zhu,
``Eikonal amplitudes on the celestial sphere,''
[arXiv:2405.15594 [hep-th]].


\bibitem{Kapec:2016jld}
D.~Kapec, P.~Mitra, A.~M.~Raclariu and A.~Strominger,
``2D Stress Tensor for 4D Gravity,''
Phys. Rev. Lett. \textbf{119}, no.12, 121601 (2017)
doi:10.1103/PhysRevLett.119.121601
[arXiv:1609.00282 [hep-th]].

\bibitem{Fotopoulos:2020bqj}
A.~Fotopoulos, S.~Stieberger, T.~R.~Taylor and B.~Zhu,
``Extended Super BMS Algebra of Celestial CFT,''
JHEP \textbf{09}, 198 (2020)
doi:10.1007/JHEP09(2020)198
[arXiv:2007.03785 [hep-th]].


\bibitem{Chen:2021xkw}
B.~Chen, R.~Liu and Y.~f.~Zheng,
``On higher-dimensional Carrollian and Galilean conformal field theories,''
SciPost Phys. \textbf{14} (2023) no.5, 088
doi:10.21468/SciPostPhys.14.5.088
[arXiv:2112.10514 [hep-th]].



\bibitem{Baiguera:2022lsw}
S.~Baiguera, G.~Oling, W.~Sybesma and B.~T.~S\o{}gaard,
``Conformal Carroll scalars with boosts,''
SciPost Phys. \textbf{14}, no.4, 086 (2023)
doi:10.21468/SciPostPhys.14.4.086
[arXiv:2207.03468 [hep-th]].

\bibitem{Rivera-Betancour:2022lkc}
D.~Rivera-Betancour and M.~Vilatte,
``Revisiting the Carrollian scalar field,''
Phys. Rev. D \textbf{106} (2022) no.8, 085004
doi:10.1103/PhysRevD.106.085004
[arXiv:2207.01647 [hep-th]].



\bibitem{deBoer:2023fnj}
J.~de Boer, J.~Hartong, N.~A.~Obers, W.~Sybesma and S.~Vandoren,
``Carroll stories,''
JHEP \textbf{09} (2023), 148
doi:10.1007/JHEP09(2023)148
[arXiv:2307.06827 [hep-th]].


\bibitem{Himwich:2020rro}
E.~Himwich, S.~A.~Narayanan, M.~Pate, N.~Paul and A.~Strominger,
``The Soft $\mathcal{S}$-Matrix in Gravity,''
JHEP \textbf{09} (2020), 129
doi:10.1007/JHEP09(2020)129
[arXiv:2005.13433 [hep-th]].

\bibitem{Pasterski:2021dqe}
S.~Pasterski, A.~Puhm and E.~Trevisani,
``Revisiting the conformally soft sector with celestial diamonds,''
JHEP \textbf{11} (2021), 143
doi:10.1007/JHEP11(2021)143
[arXiv:2105.09792 [hep-th]].

\bibitem{Freidel:2022skz}
L.~Freidel, D.~Pranzetti and A.~M.~Raclariu,
``A discrete basis for celestial holography,''
JHEP \textbf{02} (2024), 176
doi:10.1007/JHEP02(2024)176
[arXiv:2212.12469 [hep-th]].

\bibitem{Fiorucci:2023lpb}
A.~Fiorucci, D.~Grumiller and R.~Ruzziconi,
``Logarithmic celestial conformal field theory,''
Phys. Rev. D \textbf{109} (2024) no.2, 02
doi:10.1103/PhysRevD.109.L021902
[arXiv:2305.08913 [hep-th]].

\bibitem{Fan:2022vbz}
W.~Fan, A.~Fotopoulos, S.~Stieberger, T.~R.~Taylor and B.~Zhu,
``Elements of celestial conformal field theory,''
JHEP \textbf{08}, 213 (2022)
doi:10.1007/JHEP08(2022)213
[arXiv:2202.08288 [hep-th]].

\bibitem{Gradshteyn:1943cpj}
I.~S.~Gradshteyn and I.~M.~Ryzhik,
``Table of Integrals, Series, and Products,''
1943,
ISBN 978-0-12-294757-5, 978-0-12-294757-5

\bibitem{Stieberger:2022zyk}
S.~Stieberger, T.~R.~Taylor and B.~Zhu,
``Celestial Liouville theory for Yang-Mills amplitudes,''
Phys. Lett. B \textbf{836}, 137588 (2023)
doi:10.1016/j.physletb.2022.137588
[arXiv:2209.02724 [hep-th]].

\bibitem{Taylor:2023bzj}
T.~R.~Taylor and B.~Zhu,
``Celestial Supersymmetry,''
JHEP \textbf{06}, 210 (2023)
doi:10.1007/JHEP06(2023)210
[arXiv:2302.12830 [hep-th]].

\bibitem{Stieberger:2023fju}
S.~Stieberger, T.~R.~Taylor and B.~Zhu,
``Yang-Mills as a Liouville theory,''
Phys. Lett. B \textbf{846}, 138229 (2023)
doi:10.1016/j.physletb.2023.138229
[arXiv:2308.09741 [hep-th]].

\bibitem{Melton:2023lnz}
W.~Melton and S.~A.~Narayanan,
``Celestial gluon amplitudes from the outside in,''
JHEP \textbf{05}, 211 (2024)
doi:10.1007/JHEP05(2024)211
[arXiv:2312.12394 [hep-th]].

\bibitem{Giribet:2024vnk}
G.~Giribet,
``Remarks on celestial amplitudes and Liouville theory,''
[arXiv:2403.03374 [hep-th]].

\bibitem{Melton:2024gyu}
W.~Melton, A.~Sharma, A.~Strominger and T.~Wang,
``A Celestial Dual for MHV Amplitudes,''
[arXiv:2403.18896 [hep-th]].

\bibitem{Melton:2023bjw}
W.~Melton, A.~Sharma and A.~Strominger,
``Celestial Leaf Amplitudes,''
[arXiv:2312.07820 [hep-th]].

\bibitem{Huang:2023ikg}
K.~W.~Huang,
``Toward null-state equations in d \ensuremath{>} 2,''
JHEP \textbf{11}, 203 (2023)
doi:10.1007/JHEP11(2023)203
[arXiv:2308.03229 [hep-th]].

\bibitem{Huang:2024wbq}
K.~W.~Huang,
``Resummation of Multi-Stress Tensors in Higher Dimensions,''
[arXiv:2406.07458 [hep-th]].

\bibitem{Casali:2022fro}
E.~Casali, W.~Melton and A.~Strominger,
``Celestial amplitudes as AdS-Witten diagrams,''
JHEP \textbf{11}, 140 (2022)
doi:10.1007/JHEP11(2022)140
[arXiv:2204.10249 [hep-th]].

\bibitem{Fan:2022kpp}
W.~Fan, A.~Fotopoulos, S.~Stieberger, T.~R.~Taylor and B.~Zhu,
``Celestial Yang-Mills amplitudes and D = 4 conformal blocks,''
JHEP \textbf{09}, 182 (2022)
doi:10.1007/JHEP09(2022)182
[arXiv:2206.08979 [hep-th]].

\bibitem{deGioia:2022fcn}
L.~P.~de Gioia and A.~M.~Raclariu,
``Eikonal approximation in celestial CFT,''
JHEP \textbf{03}, 030 (2023)
doi:10.1007/JHEP03(2023)030
[arXiv:2206.10547 [hep-th]].

\bibitem{Gonzo:2022tjm}
R.~Gonzo, T.~McLoughlin and A.~Puhm,
``Celestial holography on Kerr-Schild backgrounds,''
JHEP \textbf{10}, 073 (2022)
doi:10.1007/JHEP10(2022)073
[arXiv:2207.13719 [hep-th]].

\bibitem{Pasterski:2020pdk}
S.~Pasterski and A.~Puhm,
``Shifting spin on the celestial sphere,''
Phys. Rev. D \textbf{104}, no.8, 086020 (2021)
doi:10.1103/PhysRevD.104.086020
[arXiv:2012.15694 [hep-th]].

\bibitem{Banerjee:2023rni}
S.~Banerjee, R.~Mandal, A.~Manu and P.~Paul,
``MHV gluon scattering in the massive scalar background and celestial OPE,''
JHEP \textbf{10}, 007 (2023)
doi:10.1007/JHEP10(2023)007
[arXiv:2302.10245 [hep-th]].

\bibitem{Ball:2023ukj}
A.~Ball, S.~De, A.~Yelleshpur Srikant and A.~Volovich,
``Scalar-graviton amplitudes and celestial holography,''
JHEP \textbf{02}, 097 (2024)
doi:10.1007/JHEP02(2024)097
[arXiv:2310.00520 [hep-th]].

\bibitem{Broedel:2013aza}
J.~Broedel, O.~Schlotterer, S.~Stieberger and T.~Terasoma,
``All order $\alpha^{\prime}$-expansion of superstring trees from the Drinfeld associator,''
Phys. Rev. D \textbf{89}, no.6, 066014 (2014)
[arXiv:1304.7304 [hep-th]].


\bibitem{Crawley:2023brz}
E.~Crawley, A.~Guevara, E.~Himwich and A.~Strominger,
``Self-dual black holes in celestial holography,''
JHEP \textbf{09}, 109 (2023)
doi:10.1007/JHEP09(2023)109
[arXiv:2302.06661 [hep-th]].

\bibitem{Costello:2022wso}
K.~Costello and N.~M.~Paquette,
``Celestial holography meets twisted holography: 4d amplitudes from chiral correlators,''
JHEP \textbf{10}, 193 (2022)
doi:10.1007/JHEP10(2022)193
[arXiv:2201.02595 [hep-th]].


\bibitem{Melton:2022fsf}
W.~Melton, S.~A.~Narayanan and A.~Strominger,
``Deforming soft algebras for gauge theory,''
JHEP \textbf{03}, 233 (2023)
doi:10.1007/JHEP03(2023)233
[arXiv:2212.08643 [hep-th]].

\bibitem{Bittleston:2023bzp}
R.~Bittleston, S.~Heuveline and D.~Skinner,
``The celestial chiral algebra of self-dual gravity on Eguchi-Hanson space,''
JHEP \textbf{09}, 008 (2023)
doi:10.1007/JHEP09(2023)008
[arXiv:2305.09451 [hep-th]].

\bibitem{Costello:2022jpg}
K.~Costello, N.~M.~Paquette and A.~Sharma,
``Top-Down Holography in an Asymptotically Flat Spacetime,''
Phys. Rev. Lett. \textbf{130}, no.6, 061602 (2023)
doi:10.1103/PhysRevLett.130.061602
[arXiv:2208.14233 [hep-th]].

\bibitem{Costello:2023hmi}
K.~Costello, N.~M.~Paquette and A.~Sharma,
``Burns space and holography,''
JHEP \textbf{10}, 174 (2023)
doi:10.1007/JHEP10(2023)174
[arXiv:2306.00940 [hep-th]].


\bibitem{Adamo:2023zeh}
T.~Adamo, W.~Bu and B.~Zhu,
``Infrared structures of scattering on self-dual radiative backgrounds,''
[arXiv:2309.01810 [hep-th]].

\bibitem{Melton:2023dee}
W.~Melton, F.~Niewinski, A.~Strominger and T.~Wang,
``Hyperbolic Vacua in Minkowski Space,''
[arXiv:2310.13663 [hep-th]].

\bibitem{Taylor:2023ajd}
T.~R.~Taylor and B.~Zhu,
``w1+\ensuremath{\infty} Algebra with a Cosmological Constant and the Celestial Sphere,''
Phys. Rev. Lett. \textbf{132}, no.22, 221602 (2024)
doi:10.1103/PhysRevLett.132.221602
[arXiv:2312.00876 [hep-th]].

\bibitem{Bittleston:2024rqe}
R.~Bittleston, G.~Bogna, S.~Heuveline, A.~Kmec, L.~Mason and D.~Skinner,
``On AdS$_4$ deformations of celestial symmetries,''
[arXiv:2403.18011 [hep-th]].

\bibitem{Adamo:2024xpc}
T.~Adamo, G.~Bogna, L.~Mason and A.~Sharma,
``Gluon scattering on the self-dual dyon,''
[arXiv:2406.09165 [hep-th]].



\end{thebibliography}
\end{document}